\documentclass[aps,preprint]{revtex4}
\usepackage[T1]{fontenc}
\usepackage[utf8]{inputenc}
\usepackage{amsmath}
\usepackage{amssymb}
\usepackage{graphicx}
\usepackage{wasysym}

\makeatletter

\providecommand{\tabularnewline}{\\}

\@ifundefined{textcolor}{}
{%
 \definecolor{BLACK}{gray}{0}
 \definecolor{WHITE}{gray}{1}
 \definecolor{RED}{rgb}{1,0,0}
 \definecolor{GREEN}{rgb}{0,1,0}
 \definecolor{BLUE}{rgb}{0,0,1}
 \definecolor{CYAN}{cmyk}{1,0,0,0}
 \definecolor{MAGENTA}{cmyk}{0,1,0,0}
 \definecolor{YELLOW}{cmyk}{0,0,1,0}
}

\usepackage{amsfonts}
\usepackage{array}
\usepackage{multirow}
\usepackage{latexsym}
\usepackage{epsfig}
\usepackage{float}
\usepackage{dsfont}

\usepackage{ragged2e}
\justifying\let\raggedright\justifying
\usepackage{caption}
\providecommand{\tabularnewline}{\\}

\newcommand{\be}{\begin{equation}}
\newcommand{\ee}{\end{equation}}

\RequirePackage{doi}

\@ifundefined{showcaptionsetup}{}{%
 \PassOptionsToPackage{caption=false}{subfig}}
\usepackage{subfig}
\makeatother

\begin{document}
\preprint{CTP-SCU/2019019}
\title{Strong cosmic censorship for a scalar field in an Einstein-Maxwell-Gauss-Bonnet-de
Sitter black hole}
\author{Qingyu Gan}
\email{gqy@stu.scu.edu.cn}

\author{Peng Wang}
\email{pengw@scu.edu.cn}

\author{Houwen Wu}
\email{iverwu@scu.edu.cn}

\author{Haitang Yang}
\email{hyanga@scu.edu.cn}

\affiliation{Center for Theoretical Physics, College of Physics, Sichuan University,
Chengdu, 610064, PR China}
\begin{abstract}
It has been shown that the Christodoulou version of the Strong Cosmic
Censorship (SCC) conjecture can be violated for a scalar field in
a near-extremal Reissner-Nordstrom-de Sitter black hole. In this paper,
we investigate the effects of higher derivative corrections to the
Einstein-Hilbert action on the validity of SCC, by considering a neutral
massless scalar perturbation in $5$- and $6$-dimensional Einstein-Maxwell-Gauss-Bonnet-de
Sitter black holes. Our numerical results exhibit that the higher
derivative term plays a different role in the $d=5$ case than it
does in the $d=6$ case. For $d=5$, the SCC violation region increases
as the strength of the higher derivative term increases. For $d=6$,
the SCC violation region first increases and then decreases as the
higher derivative correction becomes stronger, and SCC can always
be restored for a black hole with a fixed charge ratio when the higher
derivative correction is strong enough. Finally, we find that the
$C^{2}$ version of SCC is respected in the $d=6$ case, but can be
violated in some near-extremal regime in the $d=5$ case.
\end{abstract}
\maketitle
\tableofcontents{}

\bigskip{}

\section{Introduction}

It is well known that a curvature singularity could be formed during
a gravitational collapse. There might exist three types of the singularities,
namely space-like, light-like and time-like ones. Among them, the
undetermined initial data on a time-like singularity would cause the
breakdown of determinism of general relativity. Although there exist
some solutions to the Einstein's equation admitting time-like singularities
(e.g., Kerr-Newman and Reissner-Nordstrom black hole solutions), claiming
that general relativity could lose predictability is rather subtle
due to the presence of the Cauchy horizon, which encloses the time-like
singularity. In particular, to rescue the predictability of general
relativity, Penrose proposed the Strong Cosmic Censorship (SCC) conjecture,
which asserts that the maximal Cauchy development of physically acceptable
initial conditions is locally inextendible as a regular manifold \citep{Penrose:1969pc,Hawking:1969sw,Penrose:1964wq}.
Consequently, when the initial data is perturbed outside of a black
hole, whether SCC holds true crucially depends on the extensibility
of the perturbation (e.g., the metric and other fields) at the Cauchy
horizon.

To give a more rigorous definition of the extensibility of the perturbation
across the Cauchy horizon, several formulation versions of SCC have
been proposed. For example, the $C^{r}$ version of SCC states that
the metric can not be $C^{r}(r\in N)$ smooth at the Cauchy horizon
\citep{Luk:2017jxq,Luk:2017ofx}. Various evidences suggest that the
Cauchy horizon can be extendible with a continuous metric for the
perturbed initial data, indicating the falsity of the $C^{0}$ version
of SCC \citep{McNamara:1978,Dafermos:2003wr,Franzen:2014sqa}. On
the other hand, it has been argued that the $C^{2}$ version of SCC
appears to be true since the curvature blows up at the Cauchy horizon
\citep{Poisson:1990eh}. However, an observer can still experience
a finite tidal force and cross the Cauchy horizon even when the metric
is inextendible in $C^{2}$ \citep{Ori:1991zz}. So requiring that
the metric is $C^{2}$ at the Cauchy horizon seems to be too strong,
and extensions with lower smoothness shall be considered.

It is worth noting that weak solutions can have many important physical
applications, in which $C^{r}$ smooth solutions are not available.
Therefore it might be a more appropriate choice to characterize the
extensibility of the Cauchy horizon by considering whether the perturbation
is inextendible as a weak solution. For the Einstein's equation, a
weak solution is specified by locally square integrable Christoffel
symbols in some charts of the manifold. This observation then leads
to the Christodoulou formulation of SCC, which states that the maximal
Cauchy development should be inextendible as a spacetime with locally
square integrable Christoffel symbols \citep{Christodoulou:2008nj}.
Practically, for a linear scalar perturbation, the Christodoulou version
of SCC can be tested by checking whether the scalar field will belong
to the Sobolev space $H_{loc}^{1}$ at the Cauchy horizon. Note that
if the perturbation belongs to the Sobolev space $H_{loc}^{1}$, its
first derivative is locally square integrable. In other words, if
SCC is violated in the Christodoulou version, the perturbation belongs
to $H_{loc}^{1}$ and, roughly speaking, has finite energy at the
Cauchy horizon.

To check the validity of the SCC, one needs to analyze the evolution
of the perturbation, which is governed by two mechanisms. One is the
mass-inflation mechanism, associated with the exponential amplification
of a perturbation due to the blue shift effect, which might cause
a singular behavior at the Cauchy horizon \citep{Chambers:1997ef,Dafermos:2003wr,Poisson:1990eh,Ori:1991zz,Hod:1998gy,Brady:1995ni}.
The other is the time-dependent remnant perturbation decaying outside
of the black hole, which can compete with the mass inflation to invalidate
SCC. For an asymptotically flat black hole with the perturbation outside
the black hole decaying in an inverse power-law way, the mass-inflation
mechanism dominates to render the Cauchy horizon unstable \citep{Price:1971fb,Dafermos:2003wr,Dafermos:2012np,Dafermos:2014cua,Angelopoulos:2016wcv}.
Interestingly, the exponentially decay of the perturbation is observed
in asymptotically dS spacetime, indicating that mass inflation might
not be strong enough to keep SCC valid. Quantitatively, for a linear
scalar perturbation in an asymptotic dS black hole, the competition
between the the mass inflation and remnant decaying is characterized
by \citep{Costa:2014aia,Costa:2014yha,Costa:2014zha,Hintz:2015jkj,Cardoso:2017soq,Kehle:2018upl}
\begin{equation}
\beta\equiv\frac{\alpha}{\kappa_{-}},\label{eq:beta}
\end{equation}
where $\kappa_{-}$ denotes the surface gravity at the Cauchy horizon,
and $\alpha$ is the spectral gap representing the distance from real
axis to the lowest-lying QuasiNormal Mode (QNM) on the lower half
complex plane of frequency. It showed that $\beta>1/2$ corresponds
to the violation of the Christodoulou version of SCC. Moreover, $\beta>1$
represents the $C^{1}$ extensibility of a scalar field at the Cauchy
horizon, leading to the bounded curvature if coupled to gravity \citep{Costa:2014aia,Cardoso:2017soq,Dias:2018etb}.
Hence $\beta>1$ implies the violation of SCC in the $C^{2}$ version,
opening the possibility to the existence of solutions with even higher
regularity across the Cauchy horizon. From now on, the term ``SCC''
only refers to the Christodoulou version of SCC.

Recently, the validity of SCC has been widely explored in asymptotic
dS black holes. In particular, the authors in \citep{Cardoso:2017soq,Cardoso:2018nvb,Mo:2018nnu,Dias:2018ufh,Hod:2018dpx,Dafermos:2018tha,Gim:2019rkl}
considered scalar perturbations with/without mass and charge in a
Reissner-Nordstrom-de Sitter (RNdS) black hole, and found that SCC
is violated in the near-extremal region. The analysis has been extended
to the Dirac field perturbation \citep{Ge:2018vjq,Destounis:2018qnb,Zhang:2019nye,Rahman:2019uwf}
and higher space-time dimensions \citep{Rahman:2018oso,Liu:2019lon,Liu:2019rbq},
where there still exists some room for the violation of SCC. Especially
in \citep{Cardoso:2018nvb,Destounis:2018qnb}, it has been observed
that the $C^{2}$ version of SCC can be violated since $\beta>1$
appears in some near-extremal parameter regimes. Even worse, if one
considers the case with the coupled linearized electromagnetic and
gravitational perturbations in a RNdS black hole, the $C^{r}$ version
of SCC for any $r\geq2$ can be violated by taking the black hole
close enough to extremality \citep{Dias:2018etb}. Moreover, the authors
of \citep{Luna:2018jfk,Gwak:2018rba,Zhang:2019nye} argued that nonlinear
effects could not save SCC from being violated for a near-extremal
RNdS black hole. Surprisingly, SCC can always be respected for the
massless scalar field and linearized gravitational perturbations in
a Kerr-dS black hole \citep{Rahman:2018oso,Dias:2018ynt}.

It is interesting and inspiring to check the validity of SCC in models
beyond the Einstein-Maxwell theory. In \citep{Gan:2019jac,Chen:2019qbz},
we studied SCC for dS black holes in the Einstein-Born-Infeld and
Einstein-Logarithmic systems and found that the nonlinear electrodynamics
effects tend to rescue SCC. In addition, SCC has been tested for a
scalar field perturbation in the Horndeski theory in \citep{Destounis:2019omd},
which showed that the higher-order derivative couplings increases
the regularity requirements for the existence of weak solutions beyond
the Cauchy horizon. On the other hand, Gauss-Bonnet (GB) gravity,
which arises from the low energy effective action of the heterotic
string theory \citep{Boulware:1985wk}, has attracted considerable
attention in the literature. Coupling to the Maxwell electrodynamics,
namely in the Einstein-Maxwell-Gauss-Bonnet (EMGB) theory, the EMGB
black hole solution was obtained, and various aspects have been extensively
investigated \citep{Wiltshire:1985us,Cai:2001dz,Cai:2003gr,Konoplya:2004xx,Torii:2005nh,Zou:2010tv,Wang:2019urm,Zeng:2019hux}.
To the best of our knowledge, little is known about the validity of
SCC in the EMGB theory. The purpose of this paper is to investigate
the validity of SCC for a neutral massless scalar perturbation propagating
in $5$- and $6$-dimensional EMGBdS black holes.

The rest of the paper is organized as follows. In Sec.\ref{sec:EMGBdS-black-hole},
we briefly review the $5$- and $6$-dimensional EMGBdS black hole
solutions and obtain the allowed parameter regions, where the Cauchy
horizon exists. In Sec.\ref{sec:3 Quasinormal-Mode}, we show how
to compute the QNMs for a neutral massless scalar perturbation in
an EMGBdS black hole. In Sec.\ref{sec:Numerical-results}, we present
and discuss the numerical results in various parameter regions. We
summarize our results and conclude with a brief discussion in the
last section. For simplicity, we set $G=c=1$ in this paper.

\section{Einstein-Maxwell-Gauss-Bonnet-de Sitter Black Hole}

\label{sec:EMGBdS-black-hole}

In this section, we briefly review the EMGBdS black hole solution
and obtain the parameter region where three horizons exists. The action
of the Einstein-Maxwell-Gauss-Bonnet theory in $d$-dimensional spacetime
is given by \citep{Wiltshire:1985us}
\begin{equation}
S=\frac{1}{16\pi}\int d^{d}x\sqrt{-g}\left[\mathcal{R}-2\Lambda+\alpha_{GB}\left(R^{2}-4R_{\mu\nu}R^{\mu\nu}+R_{\mu\nu\rho\sigma}R^{\mu\nu\rho\sigma}\right)-F^{\mu\nu}F_{\mu\nu}\right],\label{eq:action}
\end{equation}
where $\varLambda>0$ is the cosmology constant, $\mathcal{R}$ is
the Ricci scalar curvature, and $F_{\mu\nu}=\partial_{\mu}A_{\nu}-\partial_{\nu}A_{\mu}$
is the electromagnetic tensor field of the electromagnetic field $A_{\mu}$.
It is noteworthy that the GB coupling constant $\alpha_{GB}$ is naturally
assumed to be positive since the GB correction to the Einstein gravity
is well motivated from the low energy effective action of the heterotic
string theory \citep{Boulware:1985wk}. So we focus on $\alpha_{GB}\geq0$
in this paper. In addition, the GB term is known to be topological
with no dynamics in $d=4$ dimension. Therefore, we shall consider
$d\geq5$ in what follows.

For the action (\ref{eq:action}), a static spherically symmetric
black hole solution was obtained in \citep{Wiltshire:1985us,Cai:2003gr,Zou:2010tv}:
\begin{equation}
ds^{2}=-f\left(r\right)dt^{2}+\frac{dr^{2}}{f\left(r\right)}+r^{2}d\Omega_{d-2}^{2},\quad\mathbf{A}=A_{t}dt=-\frac{4\pi Q}{(d-3)\omega_{d-2}r^{d-3}}dt,\label{eq:metric and gauge potential}
\end{equation}
with the blackening factor
\begin{equation}
f\left(r\right)=1+\frac{r^{2}}{2\widetilde{\alpha}}-\frac{r^{2}}{2\widetilde{\alpha}}\sqrt{1+4\widetilde{\alpha}\left[\frac{2\Lambda}{(d-2)(d-1)}+\frac{16\pi M}{\left(d-2\right)\omega_{d-2}r^{d-1}}-\frac{32\pi^{2}Q^{2}}{\left(d-2\right)\left(d-3\right)\omega_{d-2}^{2}r^{2d-4}}\right]},\label{eq:fr}
\end{equation}
where $M$ and $Q$ are the ADM mass and the electric charge of the
EMGBdS black hole, respectively, and $d\Omega_{d-2}^{2}$ represents
the line element of a $(d-2)$-dimensional unit sphere with volume
$\omega_{d-2}=2\pi^{(d-1)/2}/\Gamma((d-1)/2)$. For simplify, we introduce
a redefined GB parameter $\widetilde{\alpha}=\alpha_{GB}\left(d-3\right)\left(d-4\right)$.
In the limit of $\widetilde{\alpha}\rightarrow0$, eqns. (\ref{eq:metric and gauge potential})
and (\ref{eq:fr}) reduce to the $d$-dimensional RNdS black hole
as expected \citep{Liu:2019lon}.

\begin{figure}
\begin{centering}
\includegraphics[scale=0.43]{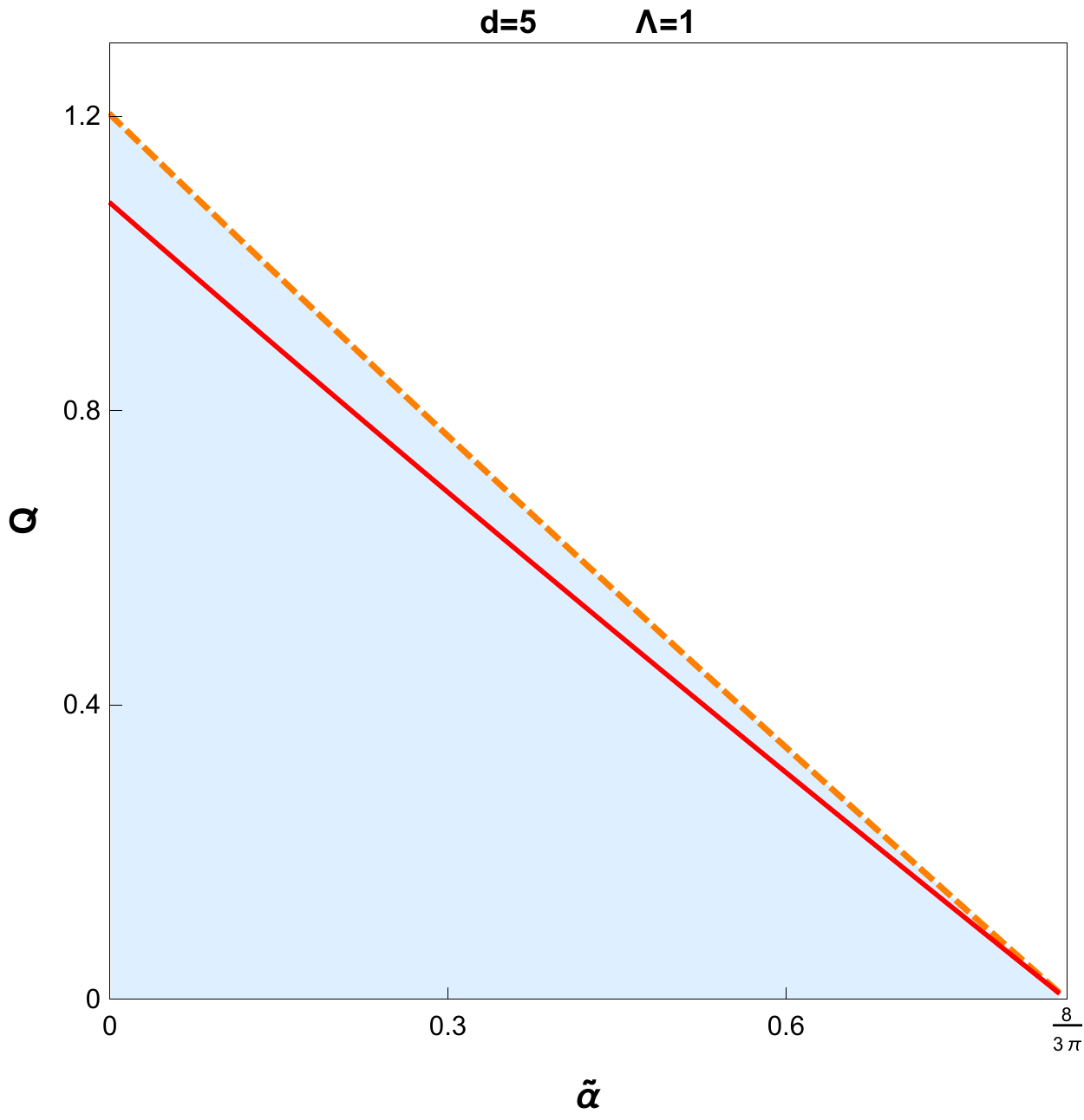}\includegraphics[scale=0.43]{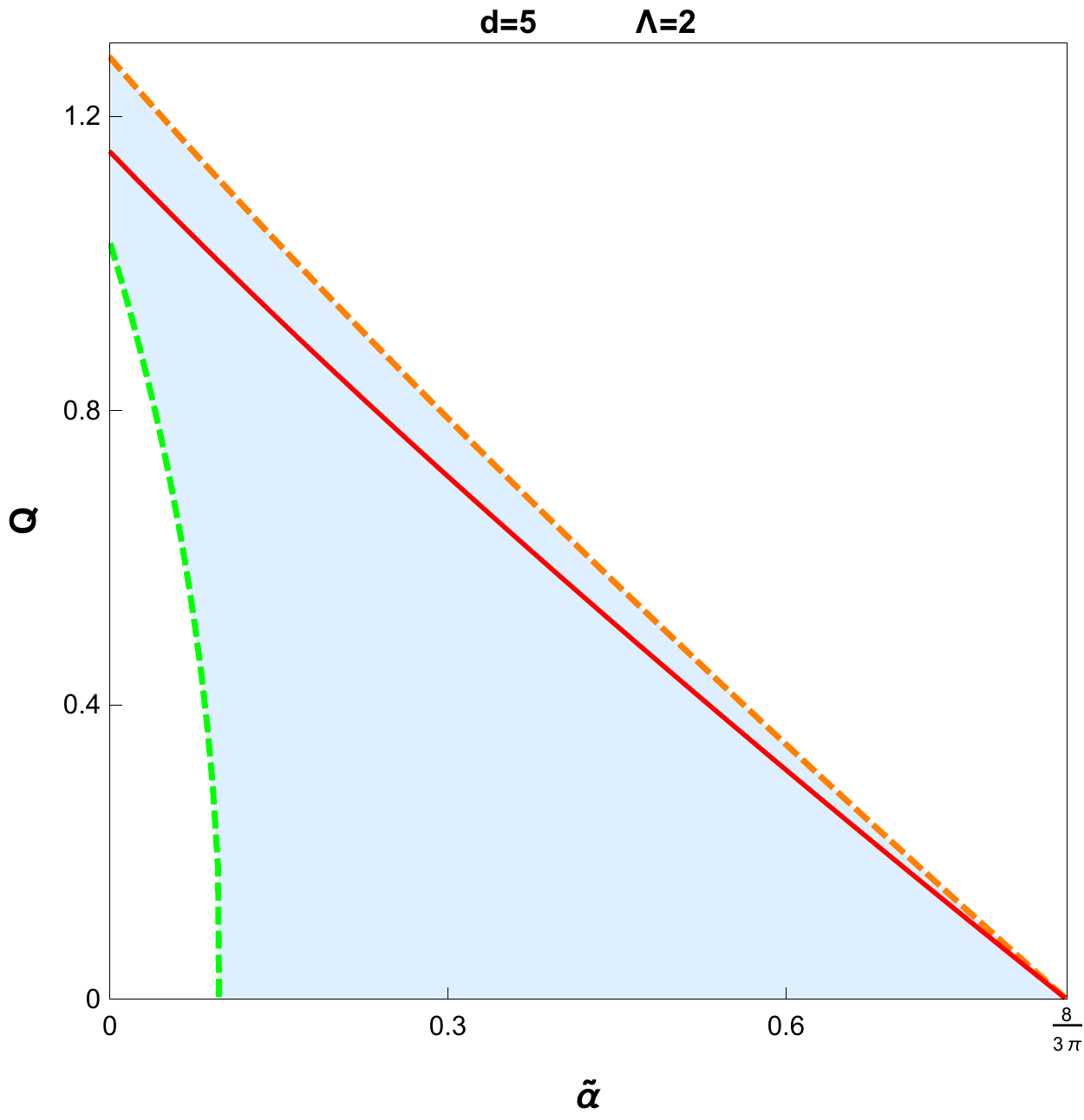}\includegraphics[scale=0.43]{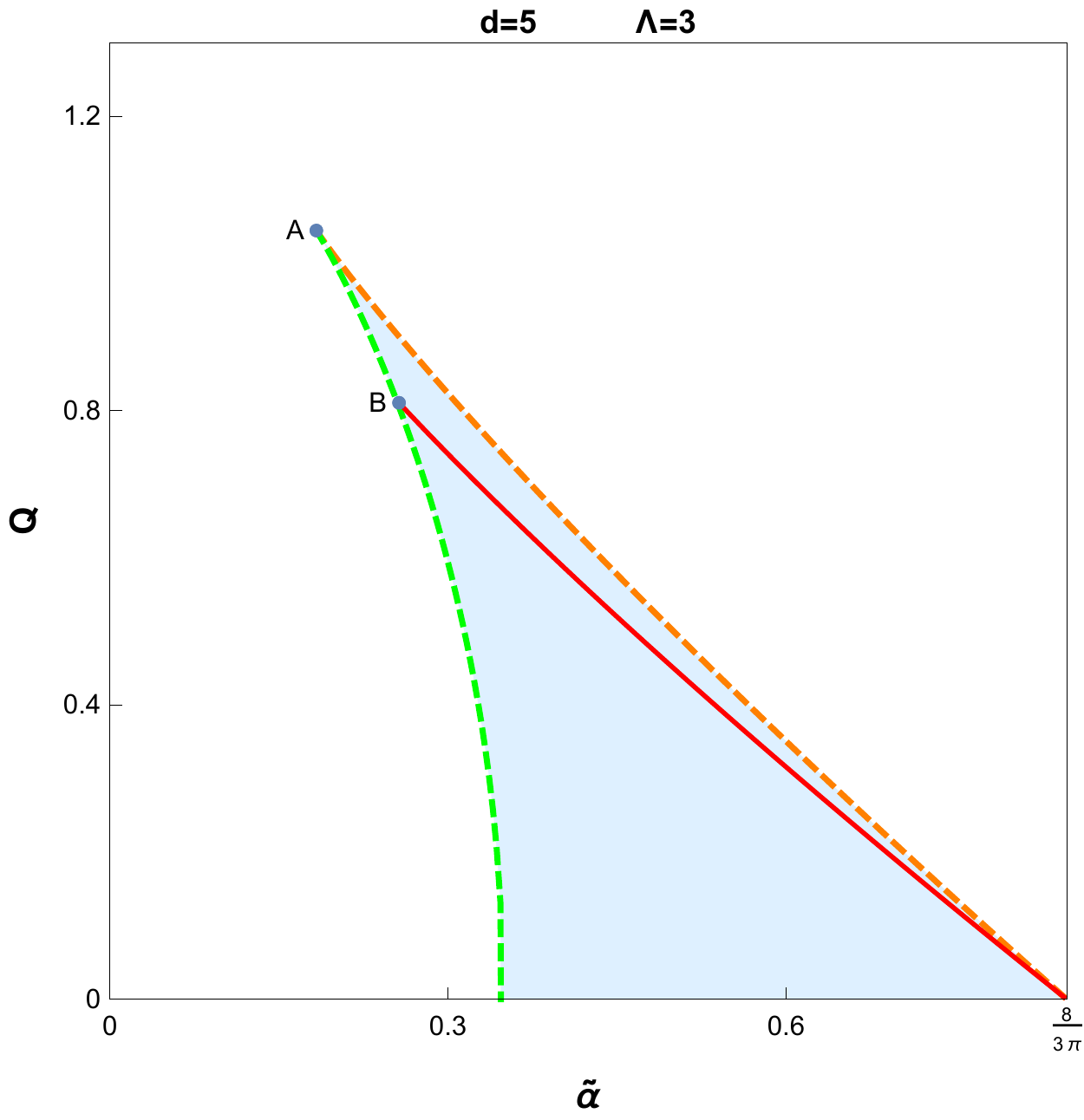}
\par\end{centering}
\begin{centering}
\includegraphics[scale=0.43]{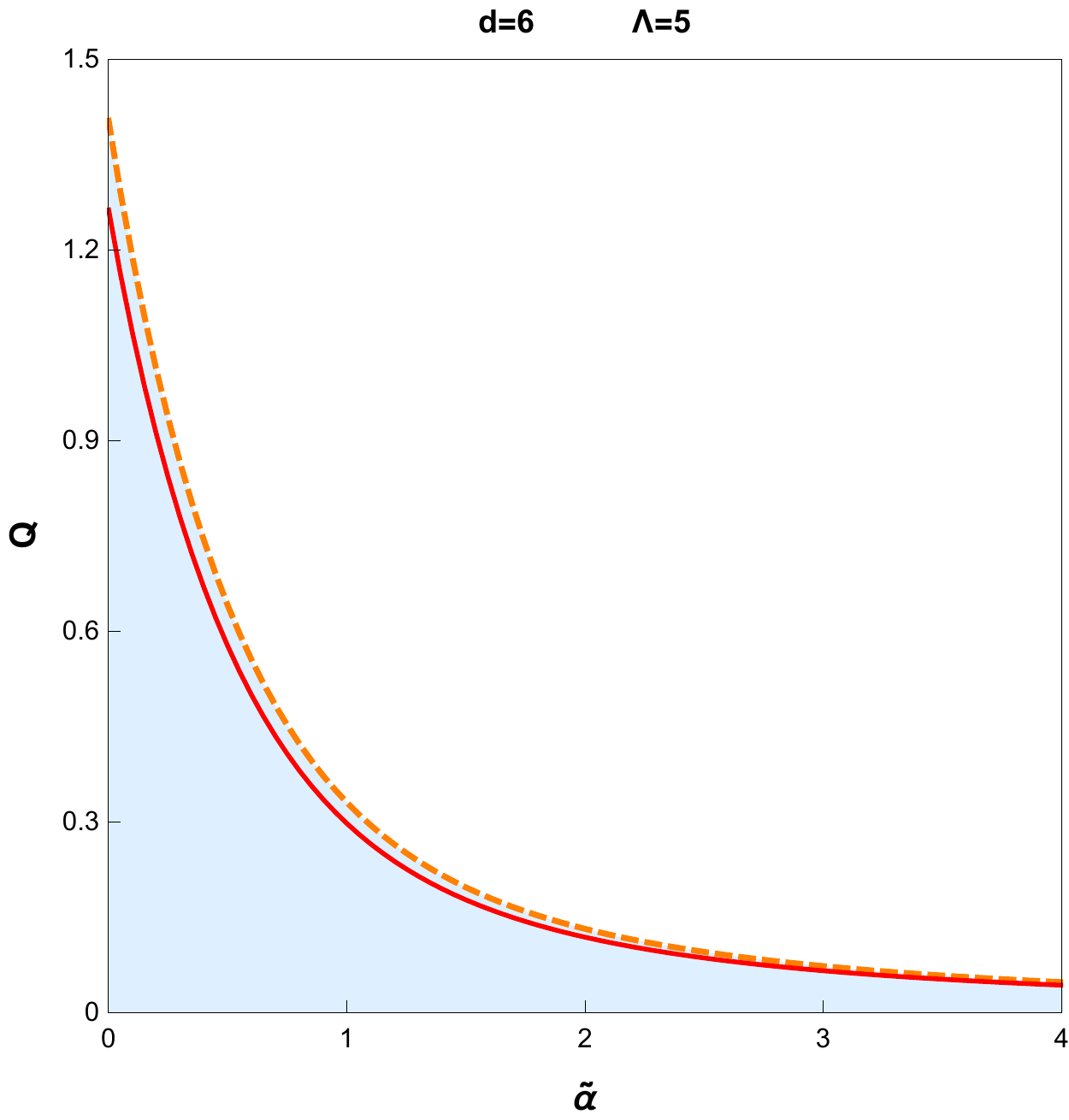}\includegraphics[scale=0.43]{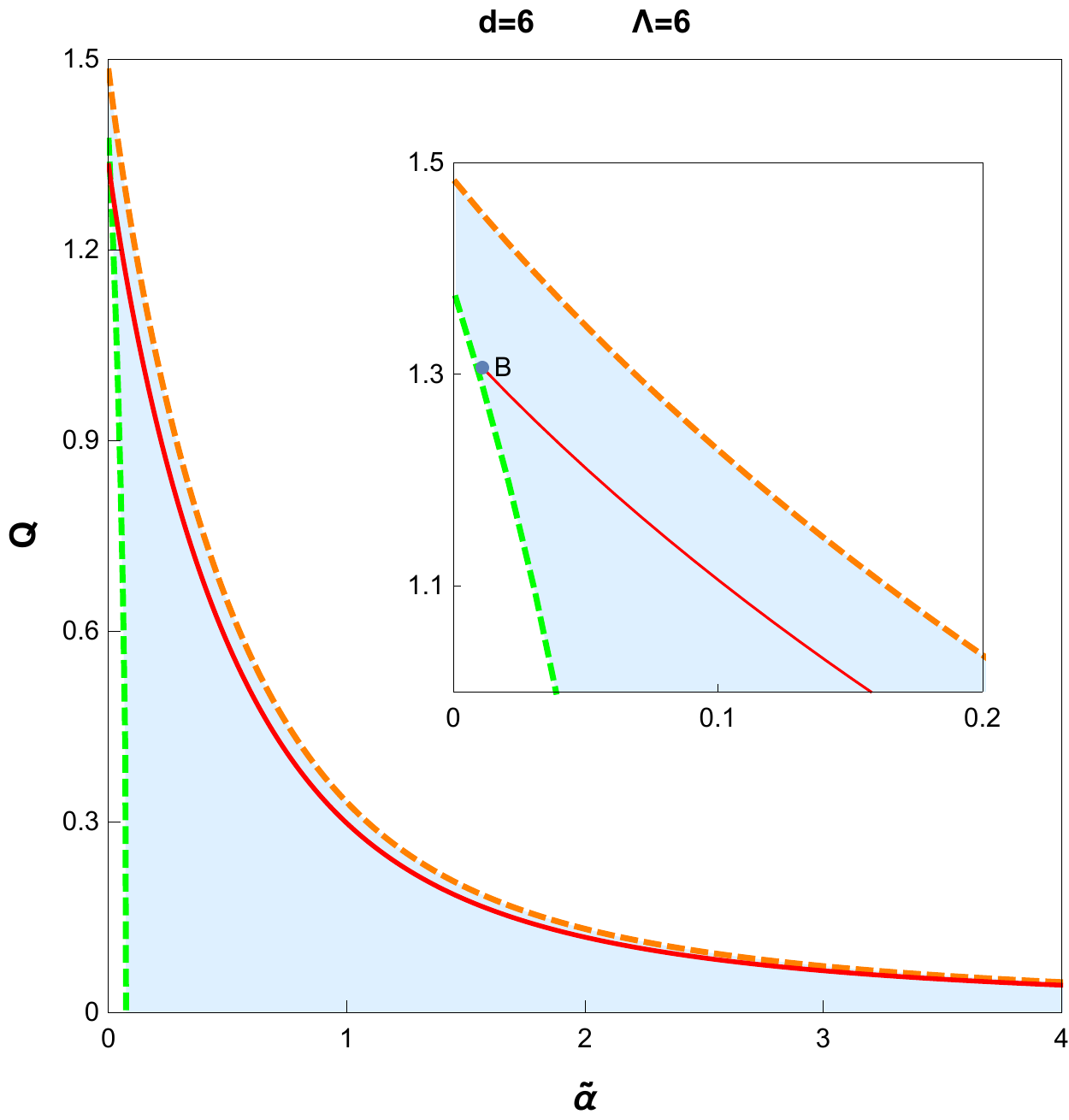}\includegraphics[scale=0.43]{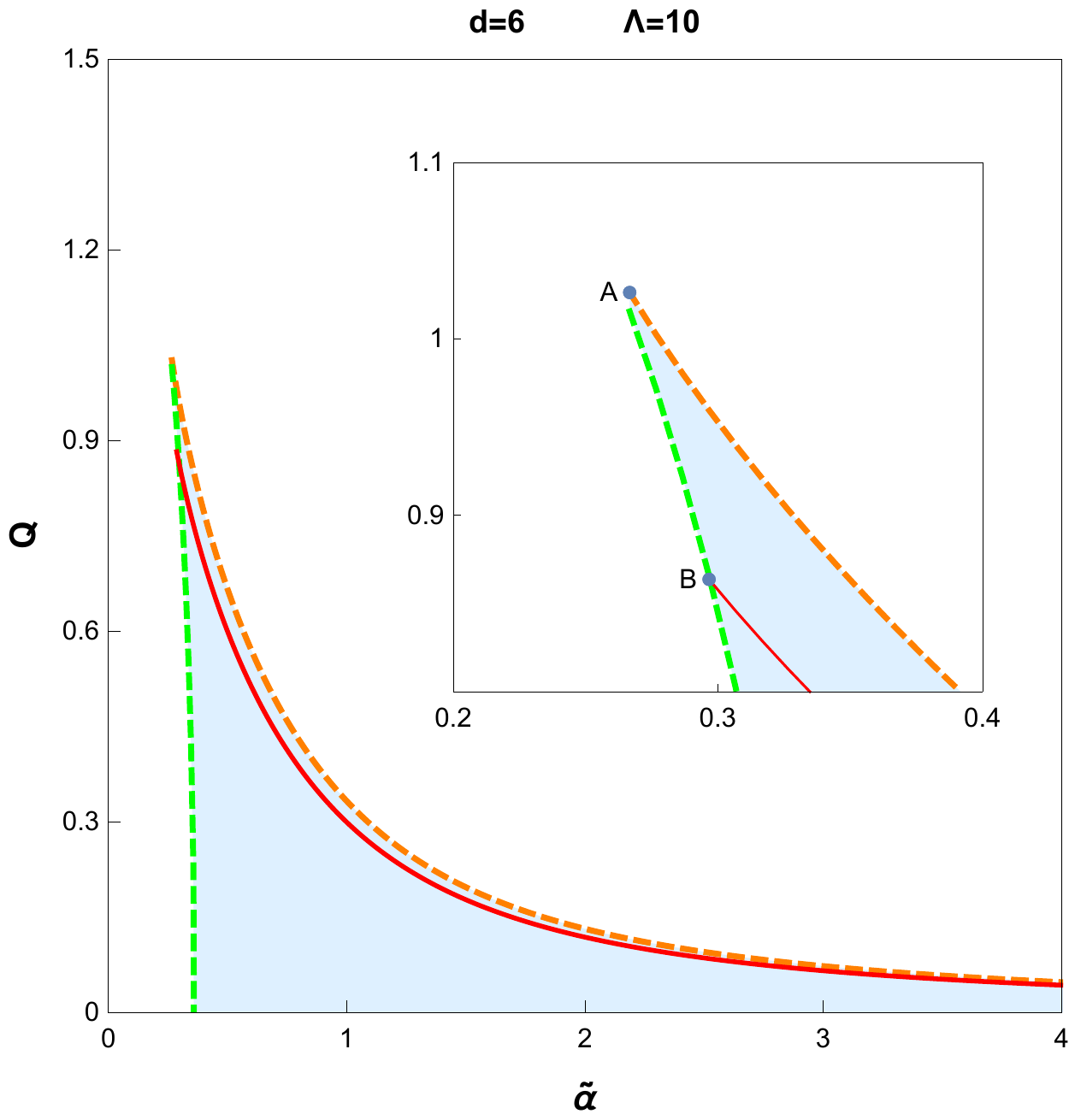}
\par\end{centering}
\begin{centering}
\includegraphics{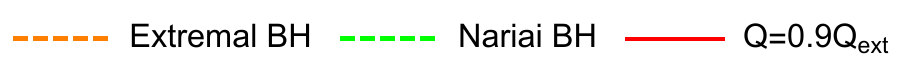}
\par\end{centering}
\centering{}\caption{{\small{}The regions in light blue are allowed to possess three horizons
for EMGBdS black holes in $d=5$ (}\textbf{\small{}upper row}{\small{})
and $d=6$ (}\textbf{\small{}lower row}{\small{}) for various values
of $\Lambda$. The dashed orange and green lines represent the extremal
black hole and the Nariai black hole, respectively. When $\Lambda>\Lambda_{\text{c}}$
(}\textbf{\small{}right column}{\small{}), there exists a tip point
$A$, which marks a nonzero minimum value of $\widetilde{\alpha}$.
The solid red line, which represents the near-extremal black hole
with the charge ratio $Q/Q_{\textrm{ext}}=0.9$, intersects with the
dashed green line at point $B$, where SCC tends to be saved as discussed
below.}}
\label{figure Region}
\end{figure}

An EMGBdS black hole is characterized by the parameters $M$, $Q$,
$\Lambda$ and $\widetilde{\alpha}$. It can show that the EMGBdS
black hole can possess one, two or three horizons in different parameter
regimes. The topology and causal structure of EMGBdS black holes have
been detailedly analyzed in \citep{Torii:2005nh}. To study SCC, we
need to find the ``allowed'' region in parameter space, in which
the black hole possesses three horizons, namely the Cauchy horizon
$r_{-}$, the event horizon $r_{+}$ and the cosmological horizon
$r_{c}$. For later use, we denote the surface gravity $\kappa_{h}\equiv\left|f^{\prime}\left(r_{h}\right)\right|/2$
with $h\in\{+,-,c\}$ for each horizon. The allowed region is determined
by the two limits, namely the extremal limit with $r_{-}=r_{+}$,
which corresponds to the extremal black hole with charge $Q_{\textrm{ext}}$,
and the Nariai limit with $r_{+}=r_{c}$, which corresponds to the
Nariai black hole with charge $Q_{\textrm{nar}}$. Hence, the allowed
region is given by $Q_{\textrm{nar}}<Q<Q_{\textrm{ext}}$ or $0<Q<Q_{\textrm{ext}}$
if no Nariai limit exists. In particular, a $5$-dimensional EMGBdS
black hole has $Q_{\textrm{nar}}=\pi\sqrt{-2+3k-2\left(1-k\right)^{3/2}}/\Lambda$
and $Q_{\text{ext}}=\pi\sqrt{-2+3k+2\left(1-k\right)^{3/2}}/\Lambda$
with $k\equiv\Lambda(8M-3\pi\widetilde{\alpha})/(6\pi)$ \citep{Thibeault:2005ha}.
The existence of $Q_{\textrm{nar}}$ and $Q_{\text{ext}}$ requires
$\widetilde{\alpha}<8M/(3\pi),$ which puts an upper bound on $\alpha.$
If $\varLambda>\Lambda_{\text{c}}\equiv3\pi/\left(4M\right)$, $Q_{\textrm{nar}}$
could only exist when $\widetilde{\alpha}>8M/(3\pi)-2/\Lambda$, which
puts a nonzero lower bound on $\alpha$. However for $d\geq6$, the
allowed region can only be numerically determined. For example, our
numerical results show that, for $d=6$, there is no upper bound on
$\widetilde{\alpha}$, and a nonzero lower bound on $\widetilde{\alpha}$
appears when $\Lambda>\Lambda_{\text{c}}\sim6.348M^{-2/3}$.

The allowed regions and their boundaries in the $\widetilde{\alpha}$-$Q$
parameter space are plotted in Fig. \ref{figure Region} for various
values of $\Lambda$ in the $d=5$ and $d=6$ cases. Without loss
of generality, we set $M=1$ in the rest of this paper. When $\Lambda>\Lambda_{\text{c}}$,
the right column of Fig. \ref{figure Region} shows that the allowed
regions have a nonzero minimum value of $\widetilde{\alpha}$, marked
by the point $A,$ as expected. One can also notice that, for the
allowed region, $\widetilde{\alpha}<8/\left(3\pi\right)$ in the $d=5$
case while no upper bound on $\widetilde{\alpha}$ exists in the $d=6$
case. Furthermore, we find that the allowed regions are in much similarity
in the $d\geq6$ cases. Thus we shall consider only two cases with
$d=5$ and $d=6$ in what follows.

\section{Quasinormal Mode}

\label{sec:3 Quasinormal-Mode}

In this section, we discuss QNMs for a neutral massless scalar perturbation
in the $d$-dimensional EMGBdS spacetime. The behavior of the neutral
massless scalar field is governed by the Klein-Gordon equation
\begin{equation}
\mathbf{\mathbf{\nabla}}^{2}\Phi=0,\label{eq:KG eq.}
\end{equation}
where $\mathbf{\nabla}$ is the covariant derivative. To facilitate
our numerical calculation, we use the Eddington-Finkelstein ingoing
coordinates $\left(v,r,\mathbf{\Omega}_{d-2}\right)$ with $v=t+r_{*}$,
where $r_{*}$ is the tortoise coordinate defined as $dr_{*}=dr/f(r)$.
In addition, we choose an appropriate gauge transformation such that
$\mathbf{A}=A_{v}dv=-4\pi Qdv/((d-3)\omega_{d-2}r^{d-3})$. Since
the EMGBdS black hole solution is static and spherically symmetric,
a mode solution of eqn. (\ref{eq:KG eq.}) can have the separable
form
\begin{equation}
\Phi(v,r,\mathbf{\Omega}_{d-2})=\mathop{\sum_{lm_{j}}}\psi_{\omega lm_{j}}(r)Y_{lm_{j}}(\mathbf{\Omega}_{d-2})e^{-i\omega v}.\label{eq:Phi}
\end{equation}
Here, $l$ and $m_{j}$ $(j=1,2,\cdots,d-3)$ denote integers required
to uniquely determine a $(d-2)$-hyperspherical harmonic $Y_{lm_{j}}(\mathbf{\Omega}_{d-2})$,
which fulfills $\nabla_{\mathbf{S}^{d-2}}^{2}Y_{lm_{j}}(\mathbf{\Omega}_{d-2})=-l(l+d-3)Y_{lm_{j}}(\mathbf{\Omega}_{d-2})$.
Since no ``magnetic splitting'' is present due to the spherical
symmetry of the background, the index $m_{j}$ can be suppressed in
$\psi_{\omega lm_{j}}$ \citep{Higuchi:1986wu,Fierro:2017fky}. Plugging
eqn. (\ref{eq:Phi}) into eqn. (\ref{eq:KG eq.}), we obtain the radial
equation
\begin{equation}
\left(r^{2}f\partial_{r}^{2}+\left(-2i\omega r^{2}+r^{2}f'+(d-2)rf\right)\partial_{r}-i(d-2)\omega r-l(l+d-3)\right)\psi_{\omega l}(r)=0,\label{eq:radial-equation}
\end{equation}
where $f^{\prime}$ denotes $df(r)/dr$.

One can perform the Frobenius method to obtain the solutions near
the event and cosmological horizons, respectively. If we impose ingoing
boundary condition at the event horizon and the outgoing boundary
condition at the cosmological horizon, namely,
\begin{equation}
\psi_{\omega l}^{\textrm{ingoing}}(r\rightarrow r_{h})\sim\textrm{const.},\qquad\psi_{\omega l}^{\textrm{outgoing}}(r\rightarrow r_{c})\sim(r-r_{c})^{-i\omega/\kappa_{c}},
\end{equation}
eqn. (\ref{eq:radial-equation}) selects a set of discrete frequencies
$\omega_{ln}$ $(n=1,2,\cdots)$, which are QNMs of the scalar field
\citep{Berti:2009kk}. There are many analytic and numerical ways
to extract QNMs \citep{Berti:2009kk,Konoplya:2011qq}. In this paper,
we employ the Chebyshev collocation scheme and the associated Mathematica
package developed in \citep{Jansen:2017oag,Jansen,url-centra.tecnico.}.
The basic idea to compute the spectrum efficiently is to discretize
the QNM equations by the pseudospectral method and solve the resulting
generalized eigenvalue equation. It can produce an additional infinite
set of purely imaginary modes, which are known to be missed by the
WKB approximation \citep{Berti:2009kk}. Moreover, WKB approximation
assumes that the potential has a single extremum, which may fail in
some cases \citep{Ishibashi:2003ap}.

To adapt to our numerical scheme in the Mathematica package, we redefine
field $\psi_{\omega l}$ as
\begin{equation}
\psi_{\omega l}=\frac{1}{x}\left(1-x\right)^{-i\omega/\kappa_{c}}\phi_{\omega l},\label{eq:redefined-phi}
\end{equation}
with a new coordinate $x\equiv(r-r_{+})/(r_{c}-r_{+})$, which renders
the new field $\phi_{\omega l}$ regular at both the event and cosmological
horizons. After the radial equation for $\phi_{\omega l}$ is obtained
from eqns. (\ref{eq:radial-equation}) and (\ref{eq:redefined-phi}),
one can use the package to find a series of QNMs, $\omega_{ln}$.
The spectral gap $\alpha$ in eqn. (\ref{eq:beta}) is then given
by $\alpha=\textrm{inf}{}_{ln}\left\{ -\textrm{Im }\omega_{ln}\right\} $.

\section{Numerical Results}

\label{sec:Numerical-results}

In this section, we present the numerical results about QNMs for a
neutral massless scalar perturbation in $5$- and $6$-dimensional
EMGBdS black holes and check the validity of SCC. These results are
obtained using the Mathematica package of \citep{Jansen:2017oag,Jansen,url-centra.tecnico.}
and found to be consistent with the results of \citep{Konoplya:2004xx,Abdalla:2005hu,Konoplya:2008ix,Liu:2019lon}
in various limits. Since it showed that SCC could be violated in a
near extremal RNdS black hole, we will focus on the near-extremal
regime of the EMGBdS black holes.

\subsection{$\mathbf{d=5}$}

\label{subsec:d5}

\begin{figure}
\begin{centering}
\includegraphics{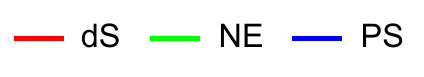}
\par\end{centering}
\begin{centering}
\subfloat[{\footnotesize{}Dominant modes of three families for various values
of $\widetilde{\alpha}$ and $\Lambda$ with varying $Q/Q_{\textrm{ext}}$.}
{\footnotesize{}The SCC violation range of $Q/Q_{\textrm{ext}}$ expands
as $\widetilde{\alpha}$ increases.}]{\begin{centering}
\includegraphics[scale=0.36]{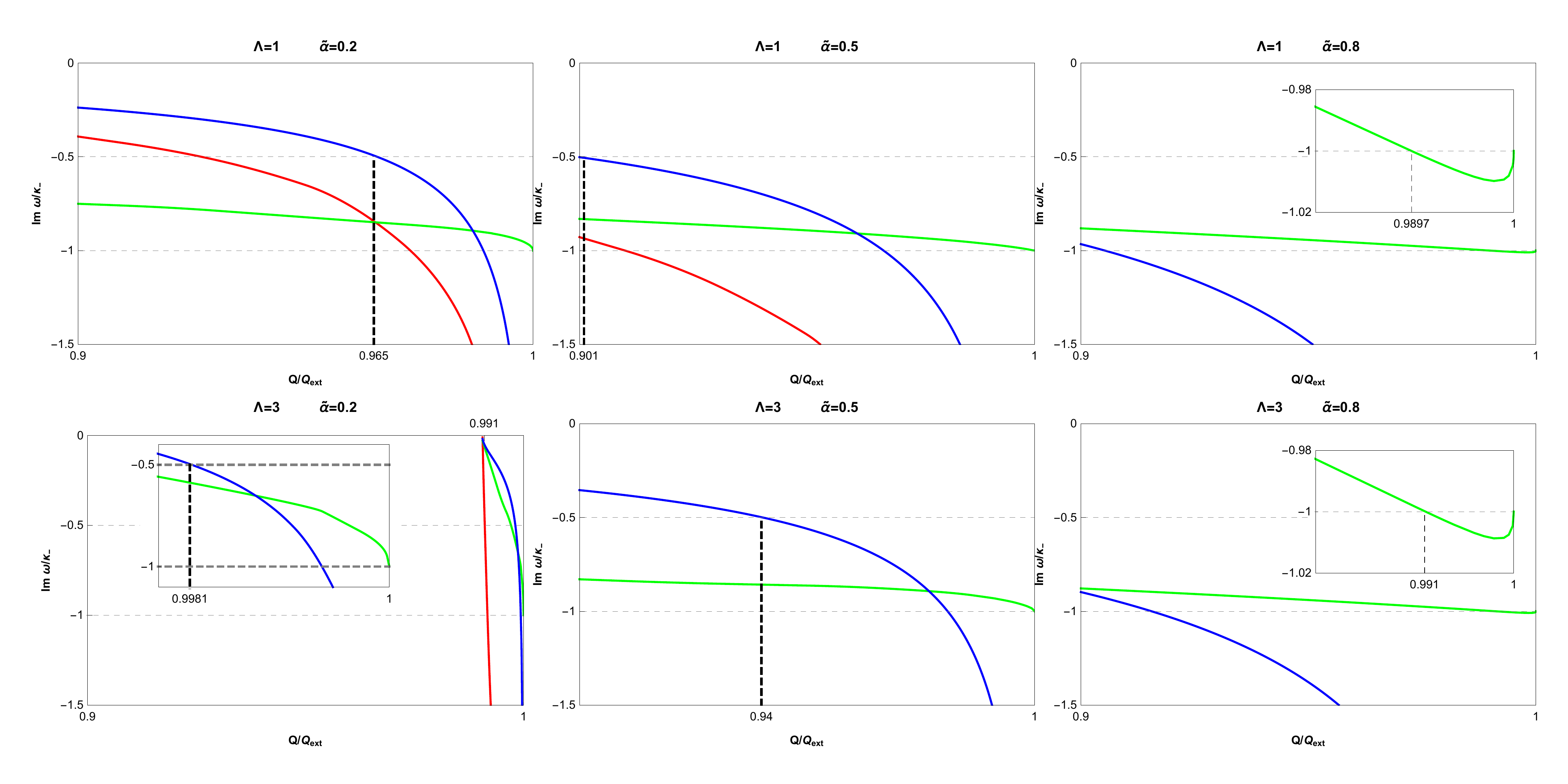}
\par\end{centering}

\centering{}\label{subfigure-d5-three-fam-fix-a}}
\par\end{centering}
\begin{centering}
\subfloat[{\footnotesize{}Dominant modes of three families for various values
of $Q/Q_{\textrm{ext}}$ and $\Lambda$ with varying $\widetilde{\alpha}$.
SCC is violated when $\widetilde{\alpha}$ is large enough.}]{\begin{centering}
\includegraphics[scale=0.36]{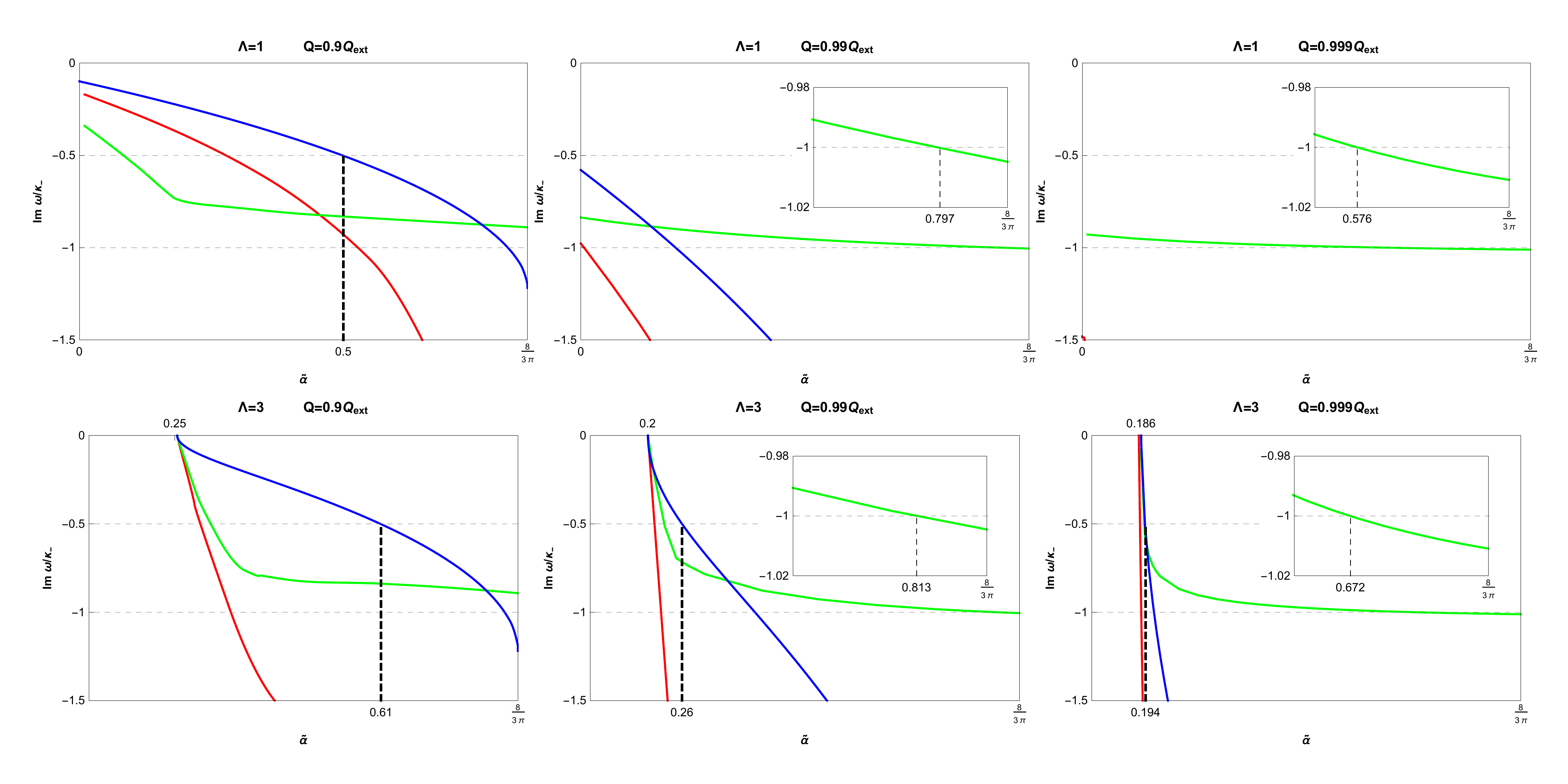}
\par\end{centering}

\centering{}\label{subfigure-d5-three-fam-fix-Q}}
\par\end{centering}
\caption{{\small{}Dominant modes of three families for a neutral massless scalar
field in a $5$-dimensional EMGBdS black hole, showing the dominant
NE mode (green lines) at $l=0$, the dominant dS mode (red lines)
at $l=1$ and the (nearly) dominant complex PS mode (blue lines) at
$l=10$. The threshold $\beta\equiv-\textrm{Im}(\omega)/\kappa_{-}=1/2$
is designated by thick dashed vertical lines, on the right of which
SCC is violated. The dashed vertical lines in the insets denote $\beta=1$.}}

\label{figure-d5-three-family}
\end{figure}

We first study the $d=5$ case. For a near-extremal black hole, it
is well known that there exist three qualitatively different families
of QNMs: \textsl{the photon sphere (PS) family}, which can be traced
back to the photon sphere, \textsl{the de Sitter (dS) family}, which
is deformation of the pure de Sitter modes, and\textsl{ the near-extremal
(NE) family}, which only appears for near-extremal black holes \citep{Cardoso:2017soq,Liu:2019lon,Destounis:2018qnb,Gan:2019jac,Claudel:2000yi}.
We plot these three distinct families for a 5-dimensional near-extremal
EMGBdS black hole in Fig. \ref{figure-d5-three-family}, where we
consider two cases with $\Lambda=1$ and $\Lambda=3$ since their
allowed regions are quite different as depicted in Fig. \ref{figure Region}.
As shown in Fig. \ref{subfigure-d5-three-fam-fix-a}, when $Q/Q_{\textrm{ext}}$
increases toward the extremal limit, $\textrm{ Im}(\omega)/\kappa_{-}$
for the PS and dS dominant modes become divergent while NE mode takes
over to make $\beta$ finite but smaller than $-1/2$. Therefore,
like a RNdS black hole, the presence of NE mode can invalidate SCC
as long as the EMGBdS black hole lies close enough to extremality.
As one increases $\widetilde{\alpha}$ from the left column to the
right column in Fig. \ref{subfigure-d5-three-fam-fix-a}, the SCC
violation range of $Q/Q_{\textrm{ext}}$ expands, which implies that
the GB term in the action (\ref{eq:action}) tends to worsen the SCC
violation for a scalar in a 5-dimensional near-extremal EMGBdS black
hole. To better understand how the GB term affects the validity of
SCC, we plot $\textrm{Im}(\omega)/\kappa_{-}$ against $\widetilde{\alpha}$
with increasing $Q/Q_{\textrm{ext}}$ from the left column to the
right column in Fig. \ref{subfigure-d5-three-fam-fix-Q}. The SCC
violation range of $\widetilde{\alpha}$, which are on the right of
the thick dashed vertical lines, increases as $Q/Q_{\textrm{ext}}$
increases, indicating that SCC tends to be violated when the black
hole is closer to extremality. Note that there is an upper bound $\widetilde{\alpha}=8/(3\pi)$
on $\widetilde{\alpha}$ for $d=5$ as discussed before. The $\textrm{Im}(\omega)/\kappa_{-}$
of all three families' dominant modes decrease to some finite values
with $\widetilde{\alpha}$ increasing toward the upper bound. Fig.
\ref{subfigure-d5-three-fam-fix-Q} displays that, for a scalar in
a 5-dimensional near-extremal EMGBdS black hole, SCC is always violated
when $\widetilde{\alpha}$ is close enough to the upper bound.

When $\varLambda=3>\Lambda_{\text{c}}$, the upper right panel of
Fig. \ref{figure Region} exhibits that there are a lower bound on
$\widetilde{\alpha}$ with fixed $Q/Q_{\textrm{ext}}$ and a possible
lower bound on $Q$ with fixed $\widetilde{\alpha}$. The black hole
with the minimum value of $\widetilde{\alpha}$ or $Q$ (e.g., the
point $B$ in Fig. \ref{figure Region}) corresponds to the Nariai
limit. The WKB method gives that the width and peak of the potential
are small in the near Nariai regime, which makes QNMs vanished in
the Nariai limit \citep{Berti:2009kk,Konoplya:2011qq,Cardoso:2017soq}.
On the other hand, the surface gravity $\kappa_{-}$ at the Cauchy
horizon remains finite in the Nariai limit, which makes $\beta=0$
for the Nariai black hole. So it is expected to observe that $\beta=0$
at the minimum values of $Q/Q_{\textrm{ext}}$ and $\widetilde{\alpha}$
in the lower left panel of Fig. \ref{subfigure-d5-three-fam-fix-a}
and the lower row of Fig. \ref{subfigure-d5-three-fam-fix-Q}, respectively.
Consequently, SCC is always valid around the Nariai limit. Therefore
when $\varLambda>\Lambda_{\text{c}}$, even for a 5-dimensional highly
near-extremal EMGBdS black hole, SCC is always saved as long as $\widetilde{\alpha}$
is close enough to its minimum value.

\begin{figure}
\begin{centering}
\includegraphics[scale=0.8]{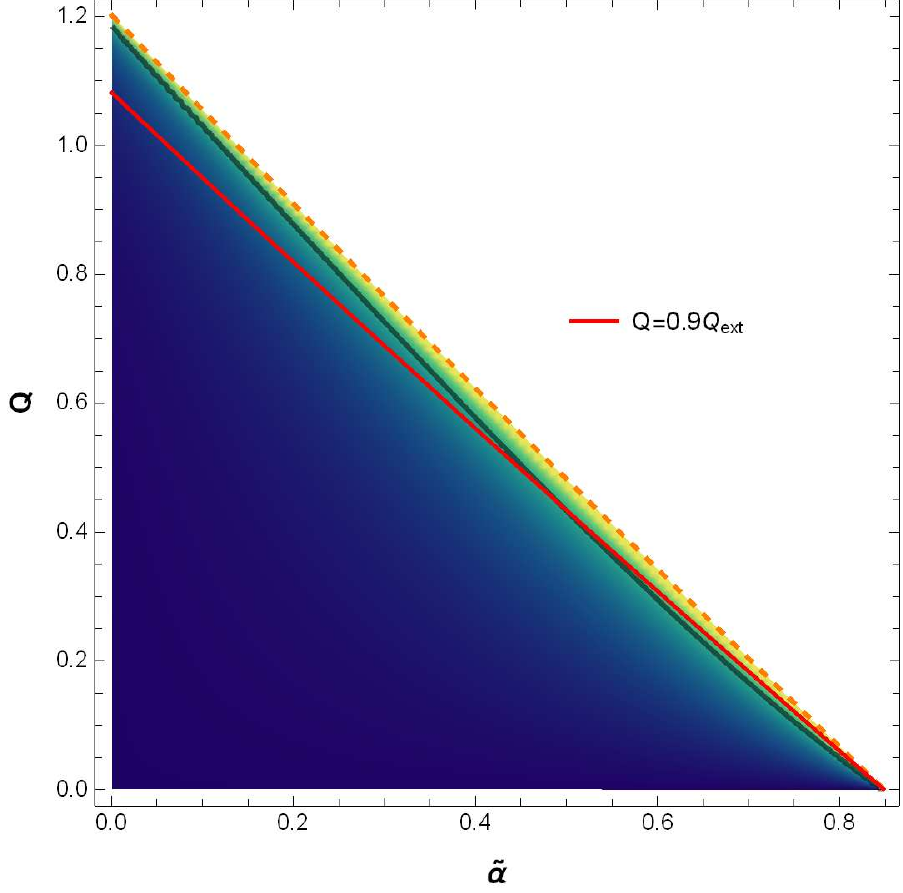}\includegraphics[scale=0.82]{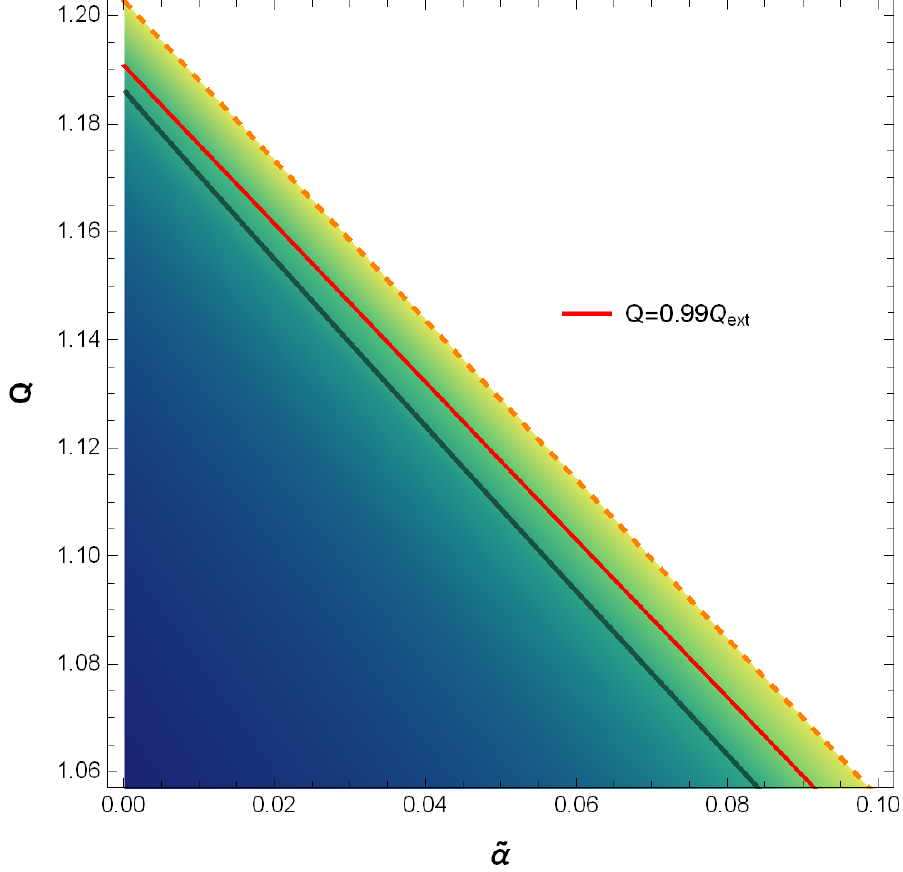}
\par\end{centering}
\begin{centering}
\includegraphics[scale=0.8]{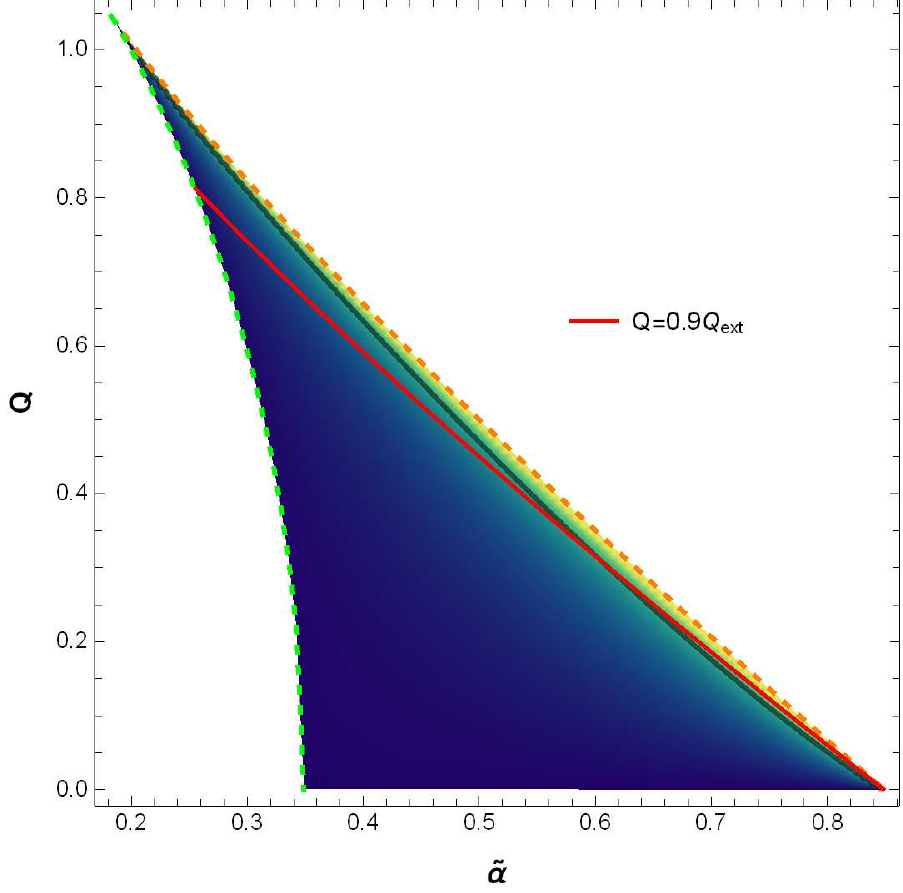}\includegraphics[scale=0.83]{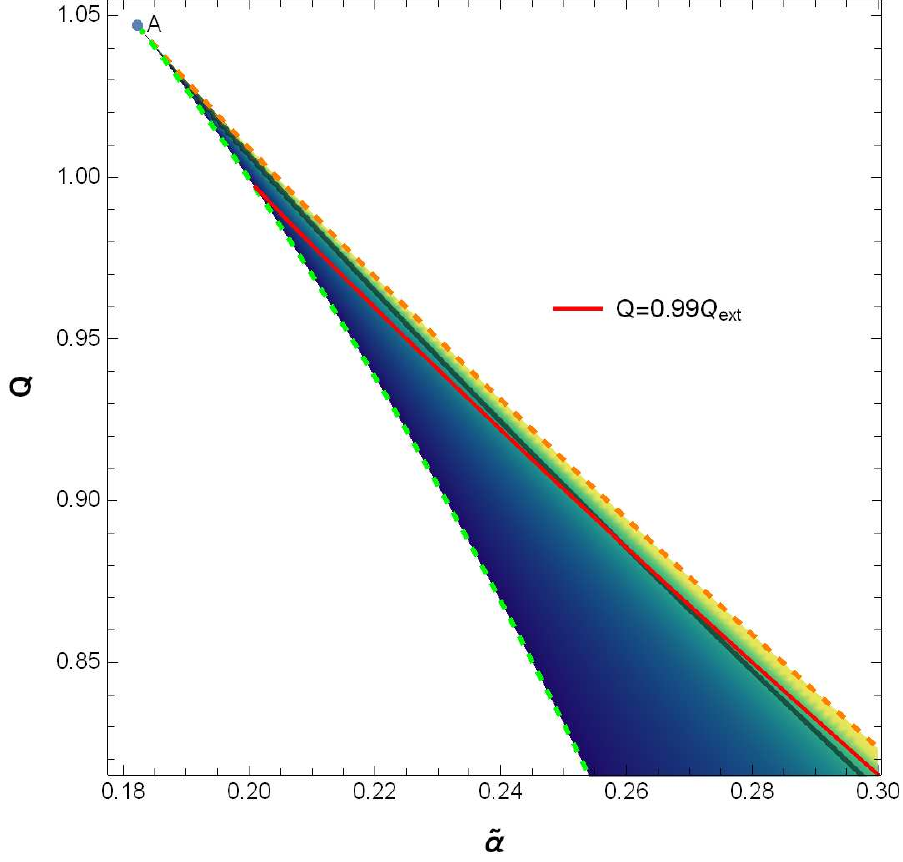}
\par\end{centering}
\begin{centering}
\includegraphics{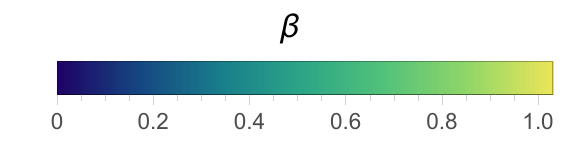}$\qquad$\includegraphics{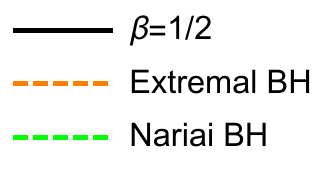}
\par\end{centering}
\caption{{\small{}Density plots of $\beta$ for a neutral massless scalar field
in a 5-dimesinal EMGBdS black hole with $\varLambda=1$ (}\textbf{\small{}upper
row}{\small{}) and $\varLambda=3$ (}\textbf{\small{}lower row}{\small{}).
The parameter space of interest is bounded by the Nariai limit (dashed
green lines) and the extremal limit (dashed orange lines). The solid
black lines represent the threshold $\beta=1/2$. SCC is valid in
the regions below the solid black lines. When $Q/Q_{\textrm{ext}}=0.9$,
SCC is saved at small $\widetilde{\alpha}$ but violated at large
$\widetilde{\alpha}$ for both $\Lambda=1$ and $\varLambda=3$. When
$Q/Q_{\textrm{ext}}=0.99$, SCC is violated in all range of $\widetilde{\alpha}$
for $\Lambda=1$ but can be saved when $\widetilde{\alpha}$ is small
enough for $\varLambda=3$.}}

\label{figure-d5-denisty-plot}
\end{figure}

We display the density plots of $\beta$ for 5-dimensional EMGBdS
black holes with $\varLambda=1$ and $\varLambda=3$ in Fig. \ref{figure-d5-denisty-plot}.
SCC is violated in the regions between the extremal lines (dashed
orange lines) and the threshold $\beta=1/2$ (solid black lines).
The $Q/Q_{\textrm{ext}}=0.9$ line in red shows that, in the both
cases with $\Lambda=1$ and $\varLambda=3$, SCC is respected at small
$\widetilde{\alpha}$ but violated when $\widetilde{\alpha}$ is large
enough. For a more extremal black hole (e.g., $Q/Q_{\textrm{ext}}=0.99$),
SCC could be always violated in the $\Lambda=1$ case. However when
$\varLambda=3$, SCC can be recovered even for a highly near-extremal
EMGBdS black hole in the region close to the Nariai line (dashed green
lines).

\begin{figure}
\begin{centering}
\includegraphics[scale=0.8]{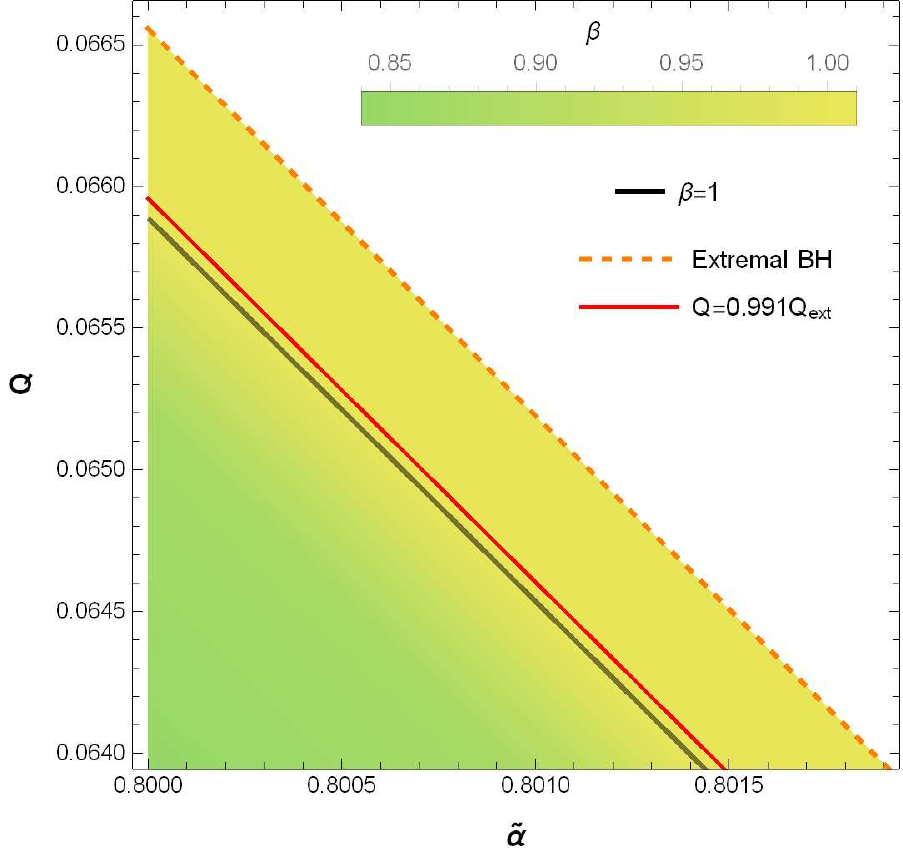}$\quad$\includegraphics[scale=0.76]{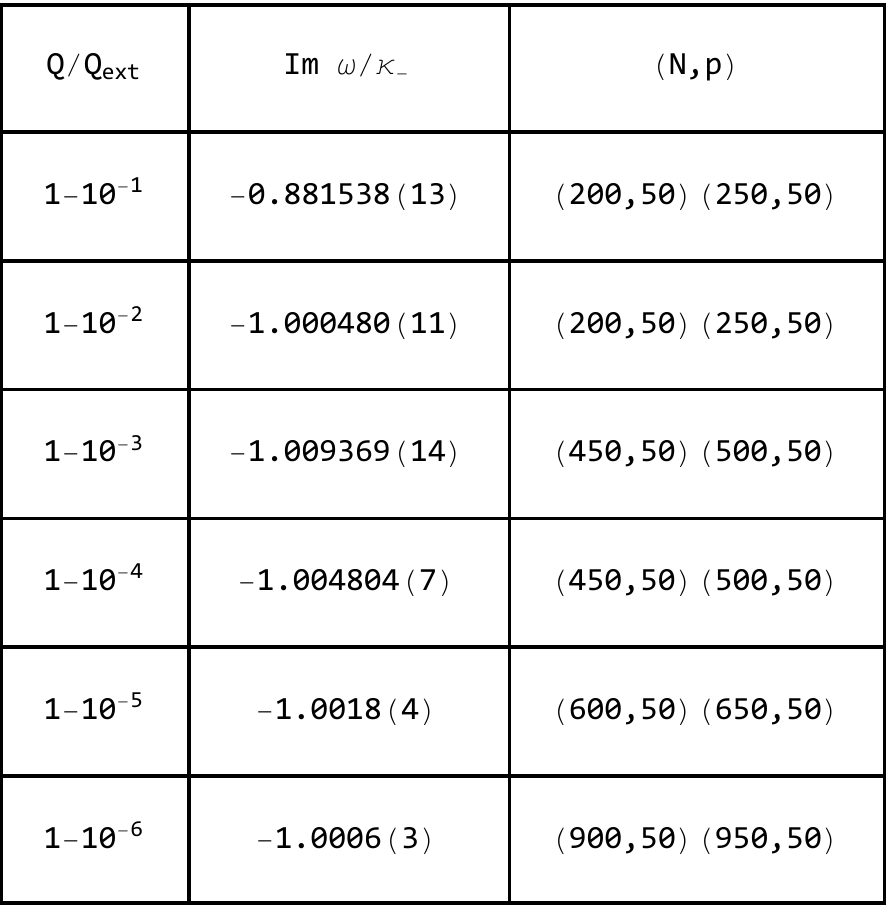}
\par\end{centering}
\caption{\textbf{\small{}Left Panel: }{\small{}Density plot of $\beta$ in
a near-extremal region around $\widetilde{\alpha}=0.8$ with $\varLambda=1$
for $d=5$. Here the solid black line corresponds to the $\beta=1$.
In the region between the solid black line and the extremal limit
(dashed orange line), the $C^{2}$ version of SCC is violated for
a near-extremal EMGBdS black hole, e.g., $Q/Q_{\textrm{ext}}=0.991$
shown in red line. }\textbf{\small{}Right Panel:}{\small{} A table
of $\textrm{Im}(\omega)/\kappa_{-}$ of the dominant NE modes for
various near-extremal values of $Q/Q_{\textrm{ext}}$ with $\varLambda=1$
and $\widetilde{\alpha}=0.8$. Numbers in brackets in the second column
indicate the number of agreed digits after the decimal point. In the
third column, we show the different grid size and precision $(N,p)$
used in the computations.}}

\label{figure-d5-beta-larger-than-1}
\end{figure}

Unlike the RNdS case \citep{Cardoso:2017soq}, the insets in Fig.
\ref{subfigure-d5-three-fam-fix-a} show that there exist some near-extremal
regions where the dominant NE mode dominates and has $\textrm{Im}(\omega)/\kappa_{-}<-1$
(i.e., $\beta>1$), which indicates the violation of SCC in the $C^{2}$
version. Moreover, our numerical results in the table of Fig. \ref{figure-d5-beta-larger-than-1}
suggest that $\textrm{Im}(\omega)/\kappa_{-}$ for the dominant NE
mode would approach $-1$ (i.e., $\beta\rightarrow1$) in the extremal
limit. In the density plot of $\beta$ displayed in Fig. \ref{figure-d5-beta-larger-than-1},
$\beta>1$ and hence the $C^{2}$ version of SCC is violated in the
region between the solid black line and the dashed orange line. Additionally,
the insets in Fig. \ref{subfigure-d5-three-fam-fix-Q} exhibit that
when $Q/Q_{\textrm{ext}}=1-10^{-2}$ and $Q/Q_{\textrm{ext}}=1-10^{-3},$
the $C^{2}$ version of SCC can be violated for a large enough value
of $\widetilde{\alpha}$, which means that GB term also tends to violate
SCC in the $C^{2}$ version.

\subsection{$\mathbf{d=6}$}

\label{subsec:d6}

\begin{figure}
\begin{centering}
\includegraphics{legendthree}
\par\end{centering}
\begin{centering}
\subfloat[{\footnotesize{}Dominant modes of three families for various values
of $\widetilde{\alpha}$ and $\Lambda$ with varying $Q/Q_{\textrm{ext}}$.
The SCC violation range of $Q/Q_{\textrm{ext}}$ first expands and
then shrinks as $\widetilde{\alpha}$ increases.}]{\includegraphics[scale=0.36]{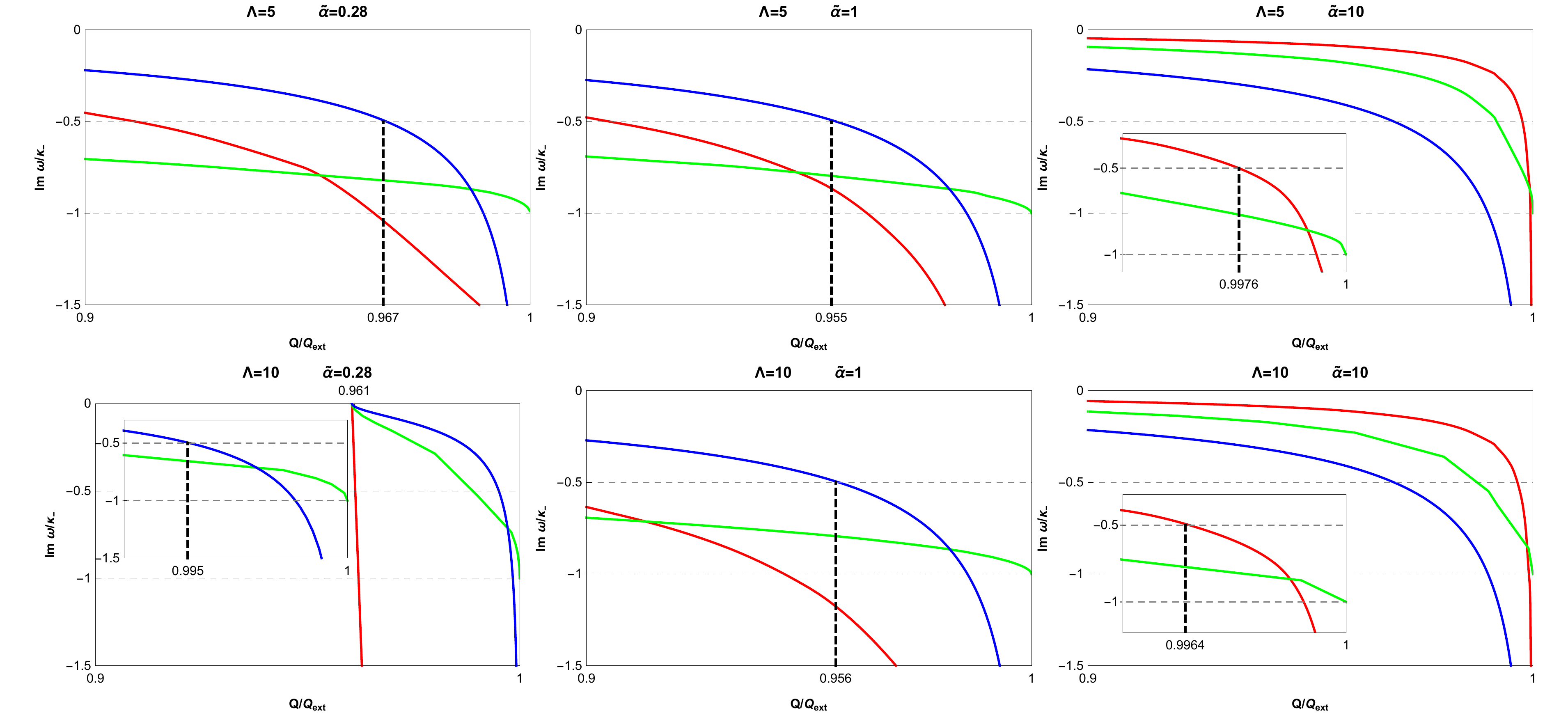}

\label{subfigure-d6-three-fam-fix-a}

}
\par\end{centering}
\begin{centering}
\subfloat[{\footnotesize{}Dominant modes of three families for various values
of $Q/Q_{\textrm{ext}}$ and $\Lambda$ with varying $\widetilde{\alpha}$.
SCC is restored when $\widetilde{\alpha}$ is large enough.}]{\includegraphics[scale=0.36]{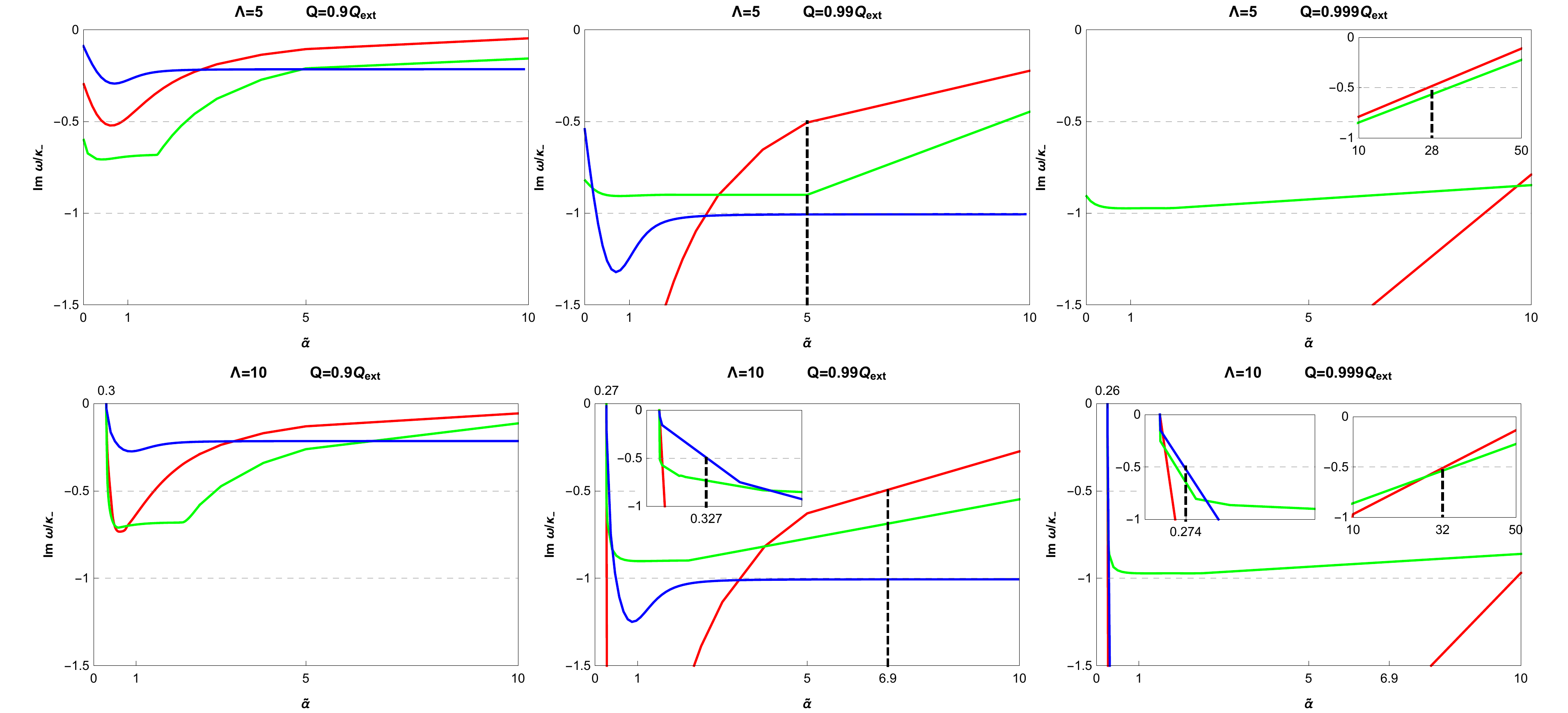}

\label{subfigure-d6-three-fam-fix-Q}

}
\par\end{centering}
\caption{{\small{}Dominant modes of three families for a neutral massless scalar
field in a $6$-dimensional EMGBdS black hole. The vertical thick
dashed lines designate the points where $\beta=1/2$.}}

\label{figure-d6-three-family}
\end{figure}

We now consider QNMs for a scalar field in a $6$-dimensional EMGBdS
black hole and investigate the validity of SCC. Since the allowed
regions in Fig. \ref{figure Region} are quite different for $\varLambda=5$
and $\varLambda=10$, we will focus on these two cases. For the lowest-lying
mode of the three families, their $\textrm{Im}(\omega)/\kappa_{-}$
are depicted against $Q/Q_{\textrm{ext}}$ for various values of $\widetilde{\alpha}$
in Fig. \ref{figure-d6-three-family}. It shows that the ranges of
$Q/Q_{\textrm{ext}}$, in which SCC is violated, first increase and
then decrease with increasing $\widetilde{\alpha}$. Furthermore,
we plot $\textrm{Im}(\omega)/\kappa_{-}$ of three families' dominant
modes against $\widetilde{\alpha}$ with fixed $Q/Q_{\textrm{ext}}$
in Fig. \ref{subfigure-d6-three-fam-fix-Q}. When $\varLambda=5<\Lambda_{\text{c}}$,
one has $\widetilde{\alpha}\geq0$. However when $\varLambda=10>\Lambda_{\text{c}}$,
there exists a positive lower bound on $\widetilde{\alpha}$, which
can be observed in the lower row of Fig. \ref{subfigure-d6-three-fam-fix-Q}.
Since the lower bound corresponds to the Nariai limit, SCC is always
valid close to the lower bound. Fig. \ref{subfigure-d6-three-fam-fix-Q}
shows that, as $\widetilde{\alpha}$ increases from its minimum value,
$\textrm{Im}(\omega)/\kappa_{-}$ of the three families' dominant
modes all first decrease, and then the dS and NE modes increase to
zero while the PS mode increases to some negative constant. Therefore
SCC is always valid as long as $\widetilde{\alpha}$ is large enough.
Moreover, the SCC violation regions in Fig. \ref{subfigure-d6-three-fam-fix-Q},
expand with increasing $Q/Q_{\textrm{ext}}$, which indicates that
SCC tends to be violated for a black hole closer to extremality.

\begin{figure}
\begin{centering}
\includegraphics[scale=0.83]{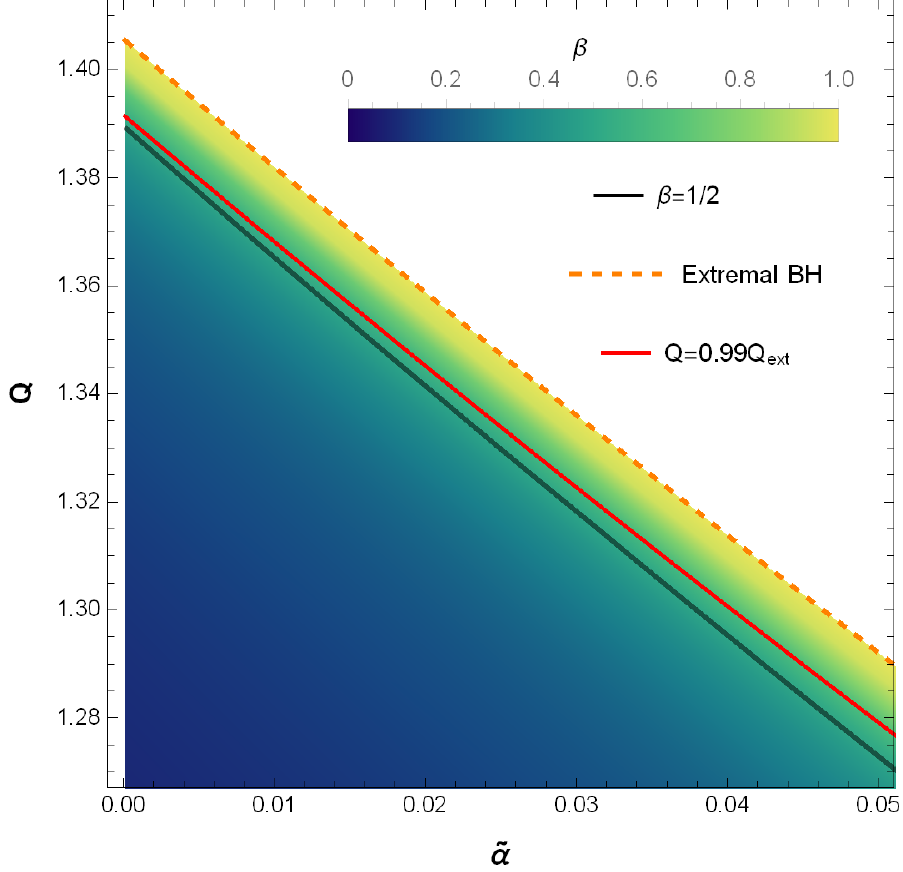}\includegraphics[scale=0.6]{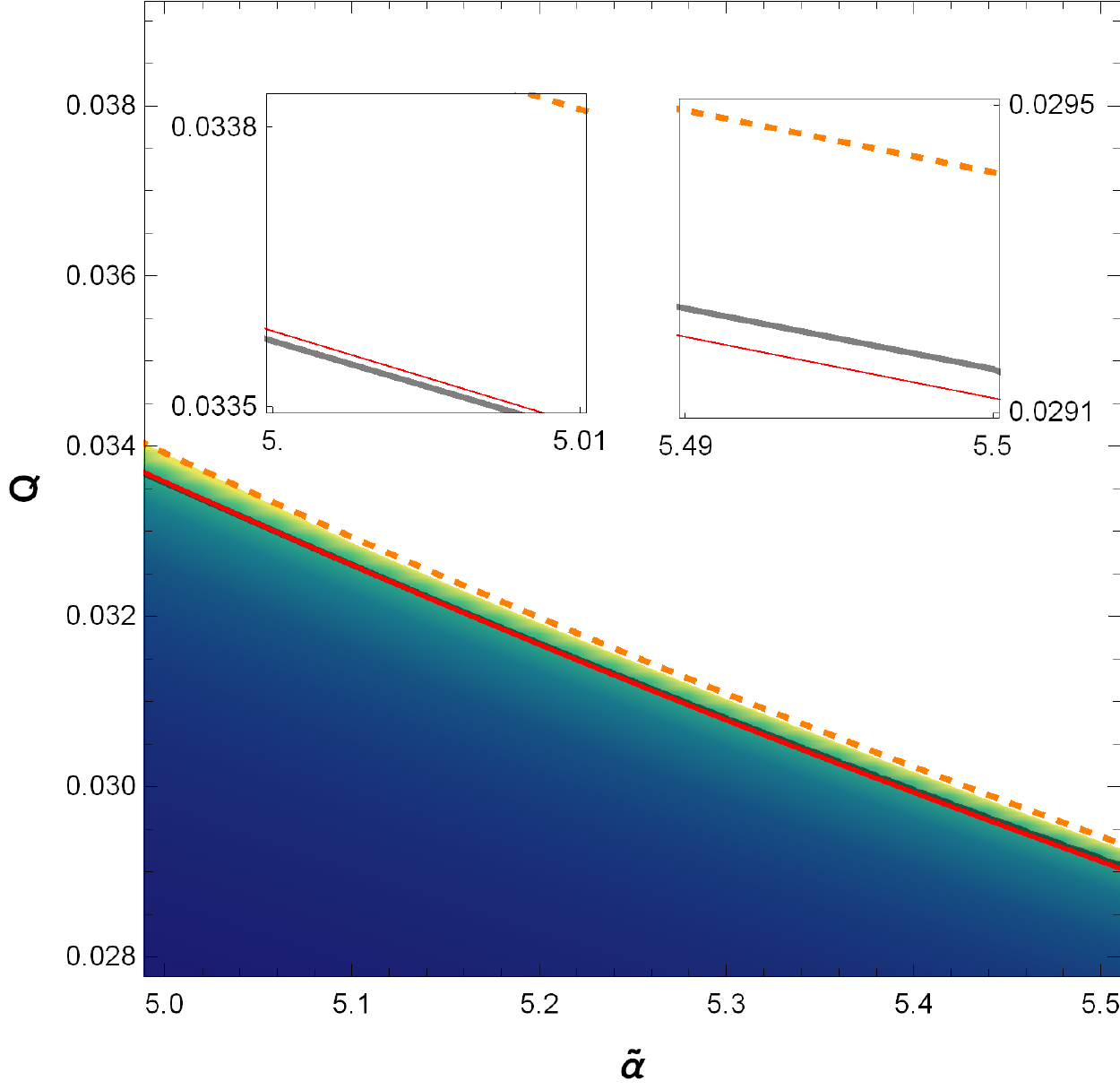}
\par\end{centering}
$\,$
\begin{centering}
\includegraphics[scale=0.83]{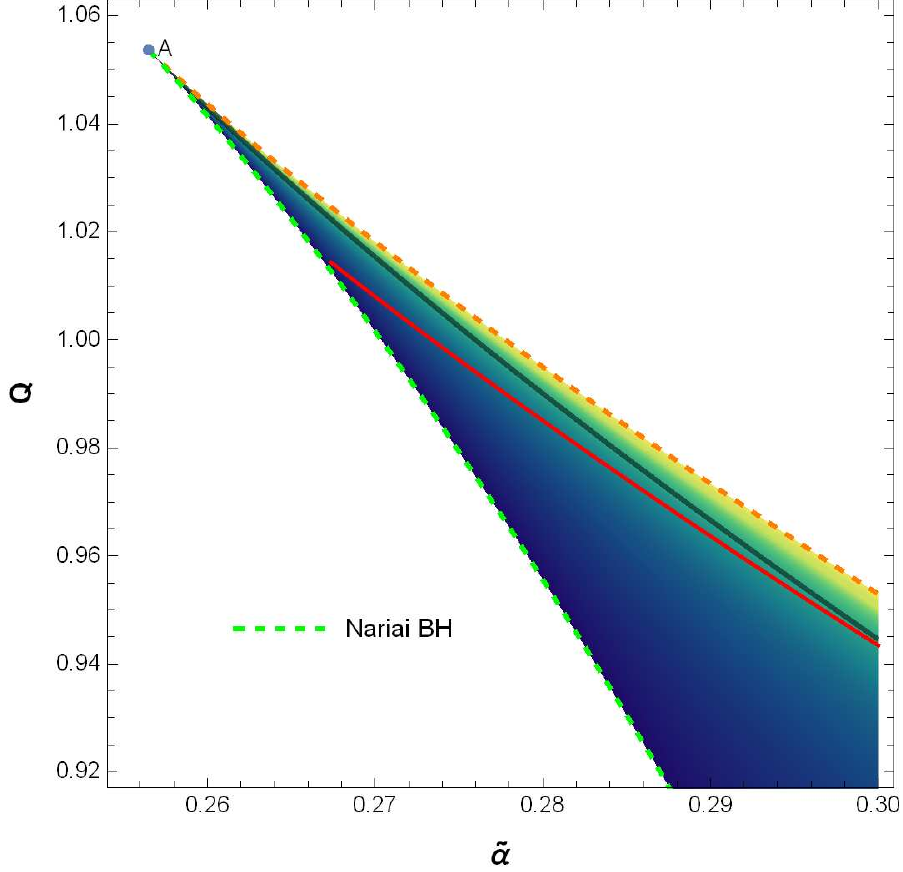}$\:$\includegraphics[scale=0.81]{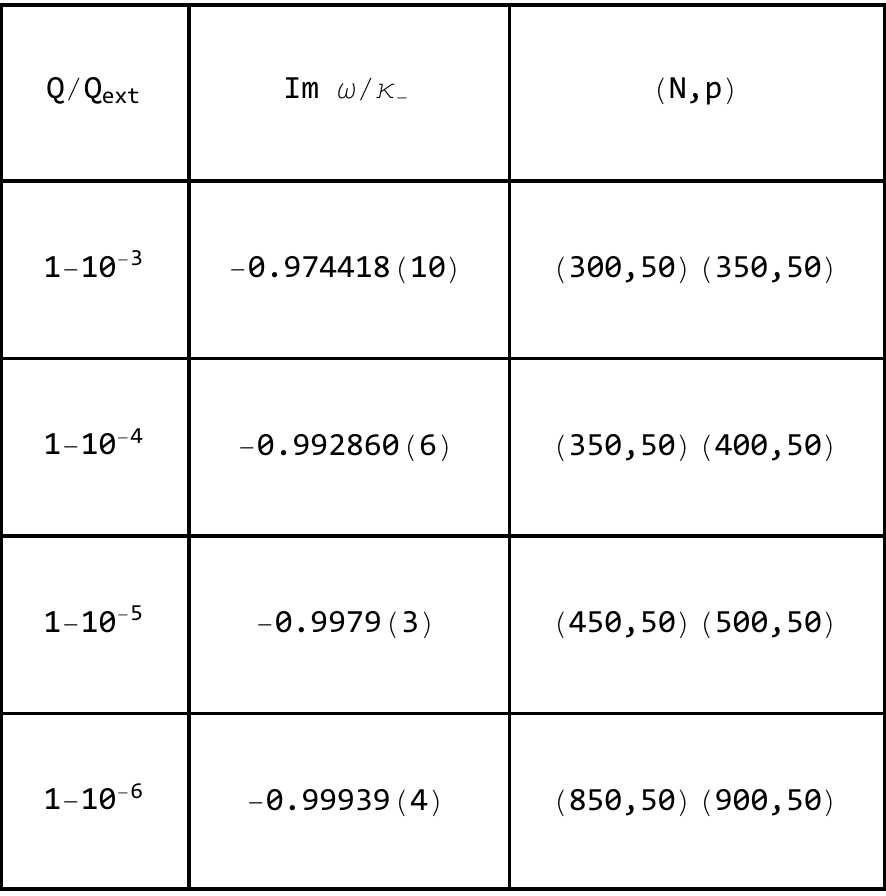}
\par\end{centering}
\caption{\textbf{\small{}Upper Row: }{\small{}Density plots of $\beta$ for
a neutral massless scalar field in a $6$-dimensional EMGBdS black
hole with $\varLambda=5$. The SCC violation regions are between $\beta=1/2$
(solid black lines) and the extremal limit (orange dashed lines).
The insets shows that the $Q/Q_{\textrm{ext}}=0.99$ line (red line)
exits the SCC violation region around $\widetilde{\alpha}\apprge5.2$.
}\textbf{\small{}Lower Left Panel: }{\small{}Density plots of $\beta$
in a $6$-dimensional EMGBdS black hole with $\varLambda=10>\Lambda_{\text{c}}$
around the minimum value of }$\widetilde{\alpha}$. {\small{}SCC is
always saved close enough to the Nariai limit (dashed green lines)}.
\textbf{\small{}Lower Right Panel:}{\small{} A table of $\textrm{Im}(\omega)/\kappa_{-}$
of the dominant NE modes for $6$-dimensional EMGBdS black holes with
$\varLambda=1$ and $\widetilde{\alpha}=1$, which suggests that $\beta<1$
in the highly near-extremal case and $\beta\rightarrow1$ in the extremal
limit.}}

\label{figure-d6-denisty-plot}
\end{figure}

The density plots of $\beta$ for 6-dimensional EMGBdS black holes
with $\varLambda=5$ and $\varLambda=10$ are displayed in Fig. \ref{figure-d6-denisty-plot},
where SCC is violated in the regions between the extremal limit (dashed
orange lines) and the threshold $\beta=1/2$ (solid black lines).
For a near-extremal black hole with $Q/Q_{\textrm{ext}}=0.99$ and
$\varLambda=5<\Lambda_{\text{c}}$ (the red line), the upper row of
Fig. \ref{figure-d6-denisty-plot} shows that SCC is violated for
small $\widetilde{\alpha}$, but can be restored when $\widetilde{\alpha}\apprge5.2$.
When $\varLambda=10>\Lambda_{\text{c}}$, the lower left panel of
Fig. \ref{figure-d6-denisty-plot} highlights the parameter space
around the lower bound on $\widetilde{\alpha}$ (point A), which shows
that SCC is always valid when black holes approach the Nariai limit.
We present $\textrm{Im}(\omega)/\kappa_{-}$ of the dominant NE modes
for various near-extremal values of $Q/Q_{\textrm{ext}}$ with $\varLambda=1$
and $\widetilde{\alpha}=1$ in the table of Fig. \ref{figure-d6-denisty-plot}.
Like the RNdS case, we observe that $\beta$ approaches $1$ from
below in the extremal limit, and hence the $C^{2}$ version of SCC
is respected in the $d=6$ case.

\section{Discussion and Conclusion}

\label{sec:Discussion-and-conclusion}

In this paper, we investigated the validity of SCC for a linear neutral
massless scalar perturbation in $5$- and $6$-dimensional EMGBdS
black holes. In Sec. \ref{sec:EMGBdS-black-hole}, we obtained the
allowed parameter regions where a EMGBdS black hole can possess the
Cauchy horizon for various $\Lambda$. After the method to calculate
QNMs was discussed in Sec. \ref{sec:3 Quasinormal-Mode}, the numerical
results were presented in Sec. \ref{sec:Numerical-results}.

For the EMGBdS black holes in the allowed region, there are two limits,
namely the extremal limit and the Nariai limit. In the extremal limit,
we numerically found that $\beta\rightarrow1$, and hence SCC is always
violated. However in the Nariai limit, we showed that $\beta\rightarrow0$,
and hence SCC is always valid. When $\varLambda>\Lambda_{\text{c}}$,
the GB parameter $\widetilde{\alpha}$ was found to have a positive
lower bound, which corresponds to the Nariai black hole. So SCC is
respected near the lower bound on $\widetilde{\alpha}$. On the other
hand, there is an upper bound on $\widetilde{\alpha}$ in the $d=5$
case while no upper bound is imposed on $\widetilde{\alpha}$ in the
$d=6$ case. Our numerical results displayed that SCC tends to be
violated/saved when $\widetilde{\alpha}$ is large enough in the $d=5$/$d=6$
case, which implies the GB term inclines to worsen/alleviate the violation
of SCC in the $d=5$/$d=6$ case. We summarize the results about how
the validity of SCC depends on $\widetilde{\alpha}$ in Table  \ref{Tab:1},
which indicates that the GB term plays a different role in the validity
of SCC for a $5$-dimensional EMGBdS black hole than it does for a
$6$-dimensional one.

\begin{table}
\begin{tabular}{|c|c|c|}
\hline
 & {\footnotesize{}SCC violation range of $Q/Q_{\textrm{ext}}$ with
fixed $\widetilde{\alpha}$} & {\footnotesize{}Varying $\widetilde{\alpha}$ with fixed $Q/Q_{\textrm{ext}}$}\tabularnewline
\hline
{\footnotesize{}$d=5$} & {\footnotesize{}increases as $\widetilde{\alpha}$ increases} & {\footnotesize{}SCC is violated near the maximum value of $\widetilde{\alpha}$}\tabularnewline
\hline
{\footnotesize{}$d=6$} & {\footnotesize{}first increases and then decreases as $\widetilde{\alpha}$
increases} & {\footnotesize{}SCC is restored when $\widetilde{\alpha}$ is large
enough}\tabularnewline
\hline
\end{tabular}\caption{{\small{}The dependence of the validity of SCC on the GB parameter
$\widetilde{\alpha}.$}}

\label{Tab:1}
\end{table}

We also checked the validity of the $C^{2}$ version of SCC for near-extremal
EMGBdS black holes. In the $d=5$ case, we found that in some parameter
region, $\beta$ is allowed to exceed unity, which implies that the
scalar is in $C^{1}$ extension on the Cauchy horizon, and hence SCC
is violated in the $C^{2}$ version. Such violation leads to the existence
of solutions with bounded Ricci curvature, corresponding to a much
more severe failure of determinism in General Relativity. However,
the violation of SCC in the $C^{2}$ version has not been observed
in the $d=6$ case. Furthermore, we numerically found that $\beta\rightarrow1$
in the extremal limit for both $d=5$ and $d=6$. To our knowledge,
the results of charged black holes in Einstein-Maxwell theory \citep{Cardoso:2017soq,Liu:2019lon},
Einstein-Born-Infeld theory \citep{Gan:2019jac}, Einstein-Logarithmic
theory \citep{Chen:2019qbz} and Horndeski theory \citep{Destounis:2019omd}
all suggest that the dominant mode of NE family approaches $1$ in
the extremal limit. It is known that such mode has been described
analytically in asymptotically flat spacetime \citep{Chen:2012zn,Kim:2012mh,Rahman:2018oso}.
The reason why $\beta\rightarrow1$ in the extremal limit may relate
to the fact that the extremal black holes share the same near horizon
topology, namely $AdS_{2}\times S^{d-2}$, leading to an enhanced
spacetime symmetry \citep{Balasubramanian:1998ee,Cvetic:1998xh,Cho:2007mn,Kunduri:2007vf,Kunduri:2008rs}.

For EMGBdS black holes with $d>6$, we expect that the result might
be similar to the $d=6$ case since their allowed regions are alike.
In this paper, we only considered the scalar field perturbation on
the fixed EMGBdS black hole background in the probe limit without
taking into account the backreaction of the scalar field on the black
hole spacetime. So our results on stability actually refer to the
scalar field rather than the EMGBdS black hole spacetime. In the future
studies, it is very interesting to check the validity of SCC and discuss
its dependence on the dimension $d$ in a full backreaction way .

\vspace{5mm}

\noindent \textbf{Note: }Just before this paper was submitted to arXiv,
a relevant preprint \citep{Mishra:2019ged} appeared, which investigated
SCC in higher curvature gravity. In \citep{Mishra:2019ged}, it was
found that the violation of SCC becomes worse as $\widetilde{\alpha}$
increases in the small $\widetilde{\alpha}$ regime for both $d=5$
and $d=6$, which is in agreement with our results. However, we carry
out the analysis in a more through way with a broader survey of the
parameter space and show that the behavior of SCC in the $d=5\text{ and }$$d=6$
cases is quite different, which was not observed in \citep{Mishra:2019ged}.
Specifically, in the large $\widetilde{\alpha}$ regime in the $d=6$
case, we display that the GB term tends to alleviate the violation
of SCC.

\vspace{5mm}

\noindent \textbf{Acknowledgements.} We thank Guangzhou Guo and Shuxuan
Ying for their helpful discussions and suggestions. This work is supported
in part by NSFC (Grant No. 11005016, 11875196 and 11375121).

\end{document}